\documentclass[11pt]{article}
\usepackage{graphicx,colortbl,booktabs,subfigure,soul}
\usepackage{amsmath, amssymb, amsthm, color, mathrsfs,calc}
\usepackage{url}
\usepackage[utf8]{inputenc}
\usepackage[english]{babel}
\usepackage{hyperref,breakurl}

\title{A stochastic SIR model for the analysis of the COVID-19 Italian epidemic}
\author{A.~Bodini, S.~Pasquali, A.~Pievatolo, F.~Ruggeri \\ CNR IMATI ``E.~Magenes'', Milano, Italy}

\date{}

\begin{document}

\maketitle

\begin{abstract}
We propose a stochastic SIR model, specified as a system of stochastic differential equations, to analyse the data of the Italian COVID-19 epidemic, taking also into account the under-detection of infected and recovered individuals in the population. We find that a correct assessment of the amount of under-detection is important to obtain reliable estimates of the critical model parameters. Moreover, a single SIR model over the whole epidemic period is unable to correctly describe the behaviour of the pandemic.
Then, the adaptation of the model in every time-interval between relevant government decrees that implement contagion mitigation measures, provides short-term predictions and a continuously updated assessment of the basic reproduction number. 
\end{abstract}


\bigskip	
\section{Introduction}
In this work we apply a stochastic version of the well-known SIR (with the three Susceptible, Infective and Removed compartments) epidemic model to the Italian COVID-19 data. The SIR model has been introduced  about ten years after the 1918 influenza pandemic, also known as Spanish flu (\cite{kermack1927contribution}), and is still popular as a simple tool to approximate disease behaviour (\cite{tolles2020modeling}). Numerous extensions such as the simpler SEIR and SIRD models or the more complex SIDARTHE (\cite{giordano2020modelling}) have been subsequently developed to make more reliable assumptions on the epidemic dynamic. To recognize the role of heterogenous contact networks in the transmission dynamics of infectious diseases, an extended SIR model with seven compartments has been developed on the nodes of a network where each node accounts for the mean number of contacts in a typical day, \cite{XUE2020108391}. For the outbreak of COVID-19, alternative models have been introduced ranging from elementary models (\cite{wang2020evaluation}) to very complex ones including spatial and temporal variations (e.g. \cite{PhysRevX.10.041055}, \cite{bai2020non}, \cite{wang2020system}, \cite{wang2020correction}). However, as a parsimonious model able to allowing the measurement and forecast of the impact of non-pharmacological interventions like social distancing, the SIR model still preserves a primary role in the analysis of the early phase of COVID-19 outbreak (\cite{bertozzi2020challenges},\cite{carletti2020covid}). In \cite{dehning2020inferring} for instance, the SIR model is combined with Bayesian parameter inference  and the effects of governmental interventions in Germany are modelled as flexible change points in the spreading rate. A SIR model with time-dependent infectivity and recovery rate which are estimated by solving an inverse problem is considered in \cite{marinov2020dynamics}.  A hierarchical epidemiological model in which two observed time series of daily proportions of infected and removed cases are emitted from the underlying infection dynamics governed by a Markov SIR infectious disease process is proposed in \cite{wangping2020extended}. By introducing a time--varying transmission rate, the Authors cover the effects of different intervention measures in dissimilar periods in Italy.
In \cite{cooper2020sir} it is assumed that the susceptible population is a
variable that can be adjusted to account for new
infected individuals spreading throughout a community. With a similar aim, \cite{9304142} include among the parameters of a SIRD model the initial number of susceptible individuals as well as the proportionality factor relating the detected number of positives with the actual (and unknown) number of infected individuals.  An alternative way to account for possible random errors in reporting is proposed in \cite{hong2020estimation} as well.

As mentioned before, several
 compartments may be added to the SIR model 
 but, in previous work (\cite{bilge2015uniqueness}) it was shown that while the parameters of the SIR model can be uniquely determined from the
time evolution of the normalized curve of removed individuals, the same does not hold for 
more complex models.
Thus, the SEIR and other models should not be used in the absence of additional information that might be obtained from clinical studies. In the present work, since we assume no clinical information, we will use the SIR model.
In particular, we introduce a stochastic SIR model to describe the dynamics of the infection, coupled with an observation equation that relates the actual numbers of infected and removed to the observed ones. There are two sources of randomness: the first in the SIR model, introduced through uncertainty in the infection and recovery rate parameters, such that they depend on time; the second in the observation equation, where a random under-reporting error is assumed. This model is amenable to sequential updating of parameters and forecasting, a useful feature when data come in on a daily basis. The updating method is a particle filter with parameter learning \cite{doucet2000sequential} for the fraction of individuals in each compartment.

The article has two objectives: on the one side, the exploration of the suitability of a SIR model, while, on the other, it provides some results on the epidemic in Italy.
The first finding is that a single SIR model is unable to capture the dynamics of the entire development of the COVID-19 epidemic, because of the numerous policy adjustments by health authorities and different government interventions that took place during the emergency. Hence we fitted one SIR model to each phase between selected government interventions, obtaining a good fit to the data. Notwithstanding, the predictive capability of this model remains very limited. The second finding is that it is very important to use the correct probability distribution for the observation error: failure to do so may produce parameter estimates that seemingly provide a good fit to the data but do not correctly describe the true underlying dynamics.

The article is organised as follows. In Section~\ref{sec:da_model} we introduce the stochastic SIR model and in Section~\ref{sec:parest} we present a discretised state-space version and the observation equation, where the state is the fraction of susceptible, infected and removed individuals. A Rao-Blackwellised particle filter (RBPF) algorithm for state filtering, 
state prediction and parameter learning is illustrated in Appendix~\ref{sec:RBPF}, \cite{doucet2000sequential}. In Section~\ref{sec:numericalsimulation} we study the filtering and prediction problem on data simulated from our model with the help of graphical displays addressing the following: the accuracy and the precision of both parameter estimation and filtering and the accuracy of the forecast (Sections~\ref{sec:datafrommodel} and \ref{sec:samplevar}); the problem of the simultaneous identifiability of the parameters in both the SIR model and the observation error distribution (Section~\ref{sec:ident}). In Section~\ref{sec:realdata} we apply our method to the Italian data of both the first and the second infection wave, obtaining a good fit, as stated above, but a forecast that may be valid only in the short term. In the same section we briefly compare the performance of a deterministic SIR model to our stochastic SIR model, showing that the latter is superior. We also consider the problem of assigning the correct observation error distribution using available information on the Infection Fatality Rate  and compare the level of the filtered state to the result of a sample serological survey carried out by Istat (Italian national statistical office) and the Italian Health Ministry between May and July 2020. This comparison indicates that our model, if correctly calibrated, provides a realistic assessment of the state of the epidemics. A section with some final remarks concludes the article.

\section{SIR model}\label{sec:da_model}

Consider a population and denote by $S_t$ the fraction of susceptible individuals at time $t$, by $I_t$ the fraction of infected individuals and by $R_t$ the fraction of recovered individuals (survivors and dead). We suppose that the population is closed, then for every time $t$
$$
S_t+I_t+R_t=1.
$$
The deterministic SIR model can be written
\begin{equation}
\label{SIRdet}
\begin{cases}
\frac{dS_t}{dt} & =-\beta I_t S_t\\
\frac{dI_t}{dt} & =\beta I_t S_t - \gamma I_t\\
\frac{dR_t}{dt} & =\gamma I_t
\end{cases}
\end{equation}
where $\beta$ is the disease transmission rate, that is, the fraction of all contacts, between infected and susceptible people, that become infectious per unit of time and per individual in the population, and $\gamma$ is the removal rate. The reciprocal of $\gamma$ is the duration of infection.
The parameters $\beta$ and $\gamma$ allow to approximate the basic reproduction number (or ratio, also called basic reproductive number or ratio)
that can be thought of as the expected number of infected people generated by an infected individual in a population where all individuals are susceptible to infection. Despite its conceptual simplicity,  $R_0$ is usually estimated with complex mathematical models developed using various sets of assumptions (\cite{delamater2019complexity}). In the above SIR model it holds
$$
R_0=\frac{\beta}{\gamma},
$$
where parameters $\beta$ and $\gamma$ are unknown and have to be estimated. We suppose that they are subject to uncertainty and change in time as follows:
\begin{equation}
\label{errore_parametri}
\beta_t=\beta_0+\sigma w^{(1)}_t \qquad \gamma_t=\gamma_0+\eta w^{(2)}_t
\end{equation}
with $w^{(1)}_t$ and $w^{(2)}_t$ independent Wiener noises, that is, $\beta_t$ is supposed normally distributed with mean $\beta_0$ and variance $\sigma^2 t$ and $\gamma_t$ is normally distributed with mean $\gamma_0$ and variance $\eta^2 t$. 
For alternative  ways to introduce stochasticity, see \cite{ganyani2021}.
The parameters $\sigma$ and $\eta$ measuring the noise intensity are assumed known and sufficiently small to obtain positive $\beta_t$ and $\gamma_t$ with probability approximately equal to one.
Substituting the expression (\ref{errore_parametri}) for $\beta$ and $\gamma$ in system (\ref{SIRdet}),
we obtain the following stochastic SIR model:
\begin{equation}
\label{SIR_stoc}
\left\{ \begin{array}{l}
dS_t=-\beta_0 I_t S_t dt -\sigma I_t S_t dw^{(1)}_t\\ 
dI_t=\left(\beta_0 I_t S_t - \gamma_0 I_t \right) dt +\sigma I_t S_t dw^{(1)}_t - \eta I_t dw^{(2)}_t \\ 
dR_t=\gamma_0 I_t dt + \eta I_t dw^{(2)}_t 
\end{array}\right.
\end{equation}

Unfortunately, the introduction of the noise in the parameters $\beta$ and $\gamma$ no longer grants the condition $S_t +I_t +R_t =1$. We can enforce it by substituting $S_t$ by $1-I_t-R_t$ in the second and third equations and removing the first equation to obtain the reduced system
\begin{equation}
\label{SIR_stoc_2eq}
\left\{ \begin{array}{l}
dI_t=\left(\beta_0 I_t (1-I_t-R_t) - \gamma_0 I_t \right) dt +\sigma I_t (1 - I_t - R_t) dw^{(1)}_t - \eta I_t dw^{(2)}_t \\ 
dR_t=\gamma_0 I_t dt + \eta I_t dw^{(2)}_t 
\end{array}\right.
\end{equation}
Denoting by $X_t=\left(I_t,R_t \right)^T$ the \textit{state} vector, by $W_t=\left(w^{(1)}_t,w^{(2)}_t \right)^T$ the vector of independent Wiener processes and by $\theta_0=\left(\beta_0,\gamma_0 \right)^T$ the parameter vector, we can rewrite system (\ref{SIR_stoc_2eq}) in vectorial form:

\begin{equation}
\label{SIR_vett}
dX_t=h\left(X_t\right)\theta_0 dt+g\left(X_t\right)  dW_t
\end{equation}
where
\begin{equation}
\label{hg}
h\left(X_t\right)=\left[\begin{array}{cc}
I_t (1-I_t-R_t) & -I_t\\
0 & I_t \\
\end{array} \right]
\qquad ; \qquad
g\left(X_t\right) =\left[\begin{array}{cc}
\sigma I_t (1-I_t-R_t) & -\eta I_t \\ 
0 & \eta I_t
\end{array}\right].
\end{equation}
We called $X_t$ the state of the system, which for COVID-19 is unobservable, and introduce $Y_t$ to denote what can be actually observed, in accordance with the terminology derived from state-space modelling. The vector $Y_t$ is characterised in the following section.

\section{Parameter estimation, filtering, forecasting, and goodness-of-fit}\label{sec:parest}

To estimate parameter $\theta_0$ we propose a Rao-Blackwellized particle filter (RBPF) algorithm based on the Euler discretization  of the stochastic system (\ref{SIR_vett}):
\begin{equation}
\label{SIR_disc}
X_{t+1}=X_t+h(X_t) \theta_0 \Delta t + g(X_t) \Delta W_t,\quad t=0, 1, 2, \ldots
\end{equation}
where we also use $t$ for discrete time to save notation. The method is described in Appendix \ref{sec:RBPF}. This algorithm allows to jointly calculate, at each time step, the estimated parameter and the state of the system using a noisy observation of the state as input. It is widely recognised that in this pandemic collected data on infected and removed suffer of under--diagnosis and  under--detection, (\cite{10665-333642}) that is, the observations are subject to an observation error. We suppose that each component of the observation vector $Y_{t+1}$ is  given by the product of the corresponding component of $X_{t+1}$ and a random variable:
\begin{equation}
\label{errore_oss}
\left[\begin{array}{c}
Y_{t,1}\\Y_{t,2}
\end{array}\right]=
\left[\begin{array}{c}
U_{t,1} X_{t,1}\\ U_{t,2} X_{t,2}
\end{array}\right]
\end{equation}
where $U_{t,1}$ and $U_{t,2}$ are independent beta distributed random variables with shape parameters $a$ and $b$. 
(In the following, by $U$, $Y$ and $X$ with no subscript we mean scalar random variables distributed as $U_{t,i}$, $Y_{t,i}$, and $X_{t,i}$, respectively, $i=1, 2$). 
The observation error in SIR models has been considered by other authors using different formulations (see, for example, \cite{stocks2020}).
Finally, we assume that the initial distribution of $\theta_0$ is Gaussian with mean $\mu_0$ and covariance matrix $\Sigma_0$.

The model can also be used to predict the future behaviour of the epidemic. Let $y_{1:t}$ be the time series of observations up to time $t$; for a fixed initial state $x_0$, the RBPF algorithm provides a sample $x_{0:t}^{(i)}$, $i=1,\ldots, M$, to approximate the posterior distribution of the state $p(x_{0:t}|y_{1:t})$. Furthermore, the conditional distributions of $\theta_0$ given  $x_{0:t}$, $p(\theta_0|x_{0:t})$, is Gaussian with mean $\mu_t=E(\theta_0|x_{0:t})$ and covariance matrix $\Sigma_t = Cov(\theta_0|x_{0:t})$. The RBPF algorithm produces a sample $(\mu_t^{(i)}, \Sigma_t^{(i)})$ of conditional mean vectors and covariance matrices given $x_{0:t}^{(i)}$.
To forecast $X_{t+k}$ given $y_{1:t}$, we aim at computing $E(X_{t+k}|y_{1:t})$. If we fix $\theta_0$ and $x_{0:t}$, and run model \eqref{SIR_disc} for $k$ time steps, we obtain a value for $X_{t+k}$ as a function $f_k(x_{0:t},\theta_0,\xi)$, in which $\xi$ indicates the sequence of increments $\Delta W_s$, $s=t+1,\ldots, t+k$. Using $f_k(x_{0:t},\theta_0,\xi)$, the conditional expectation is
\begin{equation}\label{eq:forecast}
E(X_{t+k}|y_{1:t}) = \int f_k(x_{0:t},\theta_0,\xi) p(\xi) p(\theta_0|x_{0:t}) p(x_{0:t}|y_{1:t})\, d\xi d\theta_0 dx_{0:t}
\end{equation}
where conditional independence of $\theta_0$ on $y_{1:t}$ given $x_{0:t}$ allows for substitution of $p(\theta_0|x_{0:t},y_{1:t})$ by $p(\theta_0|x_{0:t})$. Then, if for each $i$ we draw $\theta_0^{(i)}$ from $p(\theta_0|x_{0:t}^{(i)})$ and $\xi^{(i)}$ from the distribution of the Wiener process increment, the predictive expectation of $X_{t+k}$ is approximated by
\begin{equation}\label{eq:forecastMC}
	E(X_{t+k}|y_{1:t}) \simeq\frac{1}{M} \sum_{i=1}^M f_k(x_{0:t}^{(i)},\theta_0^{(i)},\xi^{(i)}) \ .
\end{equation}

To visualize how well the SIR model fits the observations, we need a way to compare the simulated trajectories with the observed data. We have supposed that the observations are smaller than the true value of the state $X$, therefore we have to scale them by a factor that makes them comparable to the filtered state.  A scaling factor is suggested by building a prediction interval of the state $X$ at each observation time. Note that from (\ref{errore_oss}) the random variable $Y/X$ is a pivotal quantity with 
beta distribution and we may state that 
\begin{equation}\label{eq:pivot}
1-q=P\left(u_{\frac q2}\leq \frac YX \leq u_{1-\frac q2}\right)=P\left(\frac Y {u_{1-\frac q2}} \leq X \leq \frac Y {u_{\frac q2}}\right)
\end{equation}
where $u_{\frac q2}$ and $u_{1-\frac q2}$ are the $\frac q2$ and the $1-\frac q2$ percentiles of the beta
distribution of $U$. Then, the corresponding prediction interval for $X$, after observing $y$, is
\begin{equation}\label{eq:cinf}
\left(
\frac y {u_{1-\frac q2}}, \frac y {u_{\frac q2}}
\right)
\end{equation}
and a natural scaling factor for a point prediction of $X$ is the median of $U$, producing $y/u_{0.5}$.
The feature of \eqref{eq:cinf} is that it does not depend on the SIR modelling assumption, but only on the observation error assumption, therefore it offers a way to see how well the SIR dynamic follows the (transformed) data.

\section{Numerical simulations}\label{sec:numericalsimulation}

\subsection{Data generated from the model}\label{sec:datafrommodel}

To check the convergence of the method we fix a parameter value and simulate the observations (or \textit{data}).

We start from an initial condition of  1\% infected and 0.1\% removed. We simulate data for the parameters $\beta_0=0.3$ and $\gamma_0=0.1$.
The parameters $\sigma$ and $\eta$ in (\ref{errore_parametri}) are 0.03 and 0.01, respectively.

\begin{figure}
\center
\includegraphics[scale=0.6]{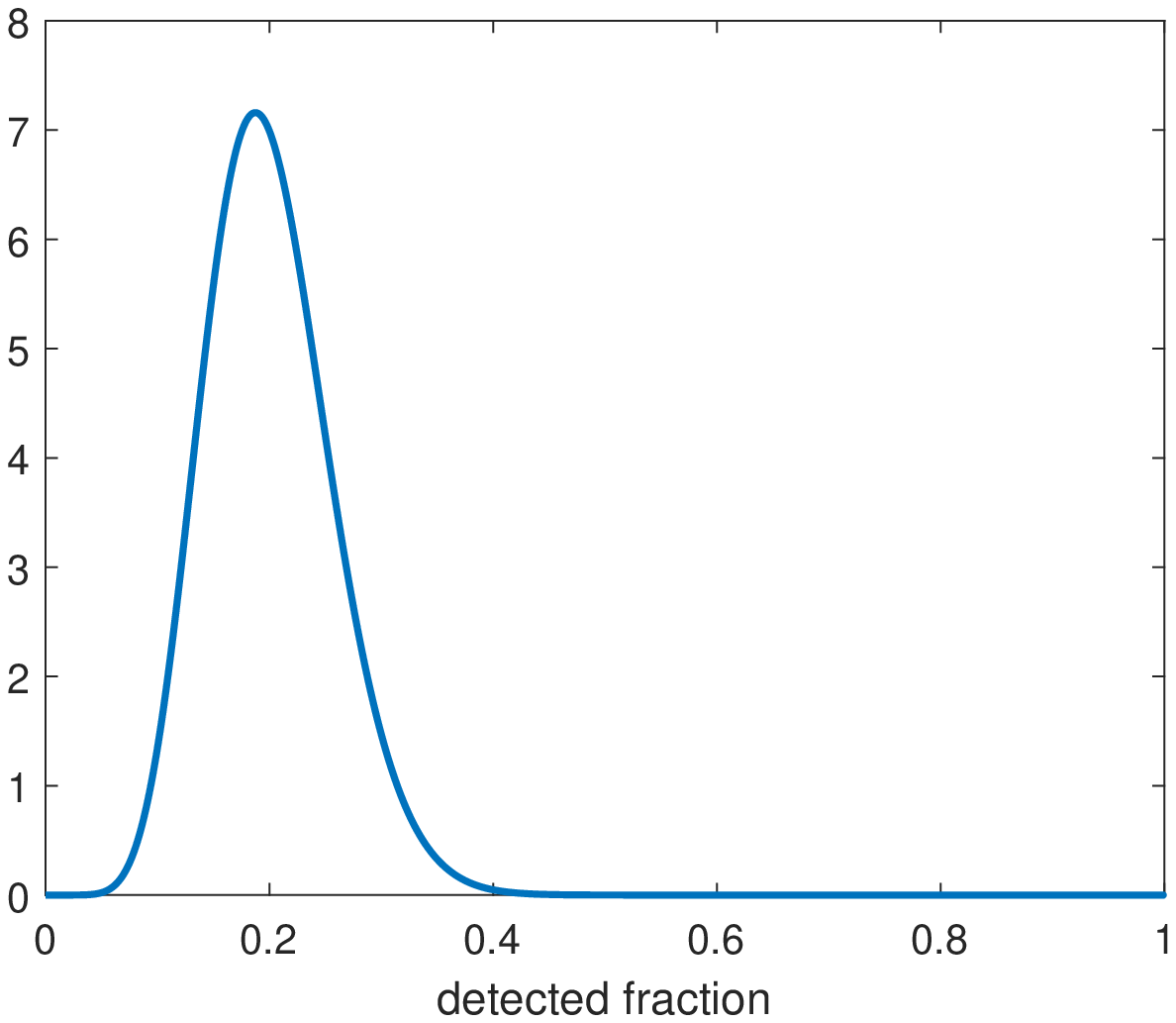}
\hspace{1cm}
\includegraphics[scale=0.6]{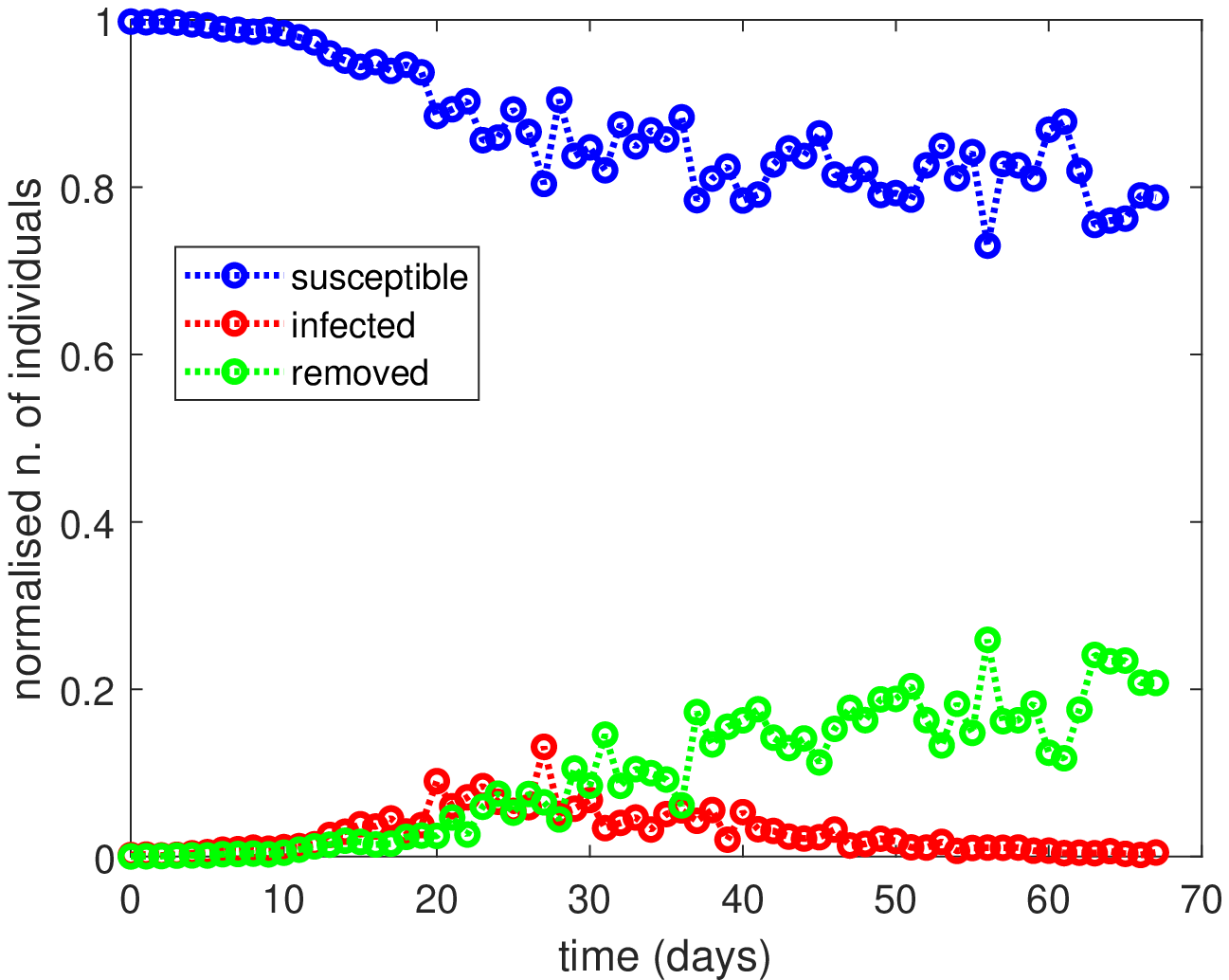}
\caption{Left panel: beta distribution with parameters $a=10$ and $b=40$ for the observation error terms $U_{t,1}$ and $U_{t,2}$ in \eqref{errore_oss}. Right panel: data generated by \eqref{errore_oss} from state trajectories generated by (\ref{SIR_stoc_2eq}).}
\label{graf_gamma_dati}
\end{figure}

We run model (\ref{SIR_stoc_2eq}) with an underlying time step of $1/24$ day to generate 67 daily step states. The first 60 states will be used for the estimation procedure and the other 7 to check the goodness of the forecast.
Then, we use equation (\ref{errore_oss}) with $a=10$ and $b=40$ \ 
for the beta 
distribution of the observation error. This distribution is reported in the left panel of Figure \ref{graf_gamma_dati} and the simulated observations are in the right panel of Figure \ref{graf_gamma_dati}.

We apply the RBPF algorithm described in Appendix \ref{sec:RBPF} with 200,000 particles and with a time step of $1/24$ day.
Since we have a single observation for each day, to run the algorithm we have to impute new hourly observations
by means of a linear interpolation between two consecutive daily observations. The imputed
observations are no longer indipendently distributed conditionally on the states, 
however this approximate procedure keeps the effective
sample size of the RBPF algorithm at large values with no appreciable difference on the results.

\begin{figure}
	\center
	\includegraphics[scale=0.6]{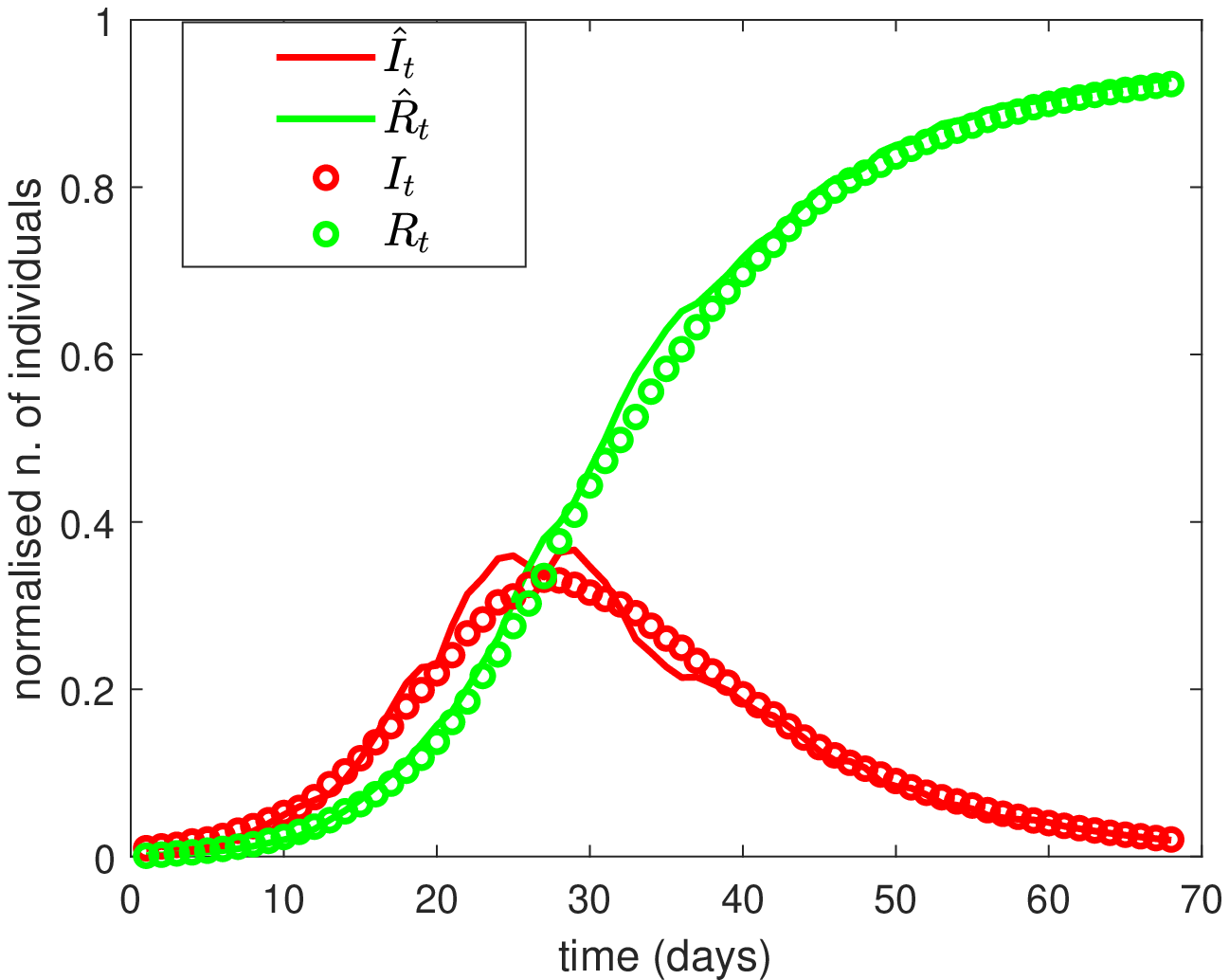}
	\hspace{1cm}
	\includegraphics[scale=0.6]{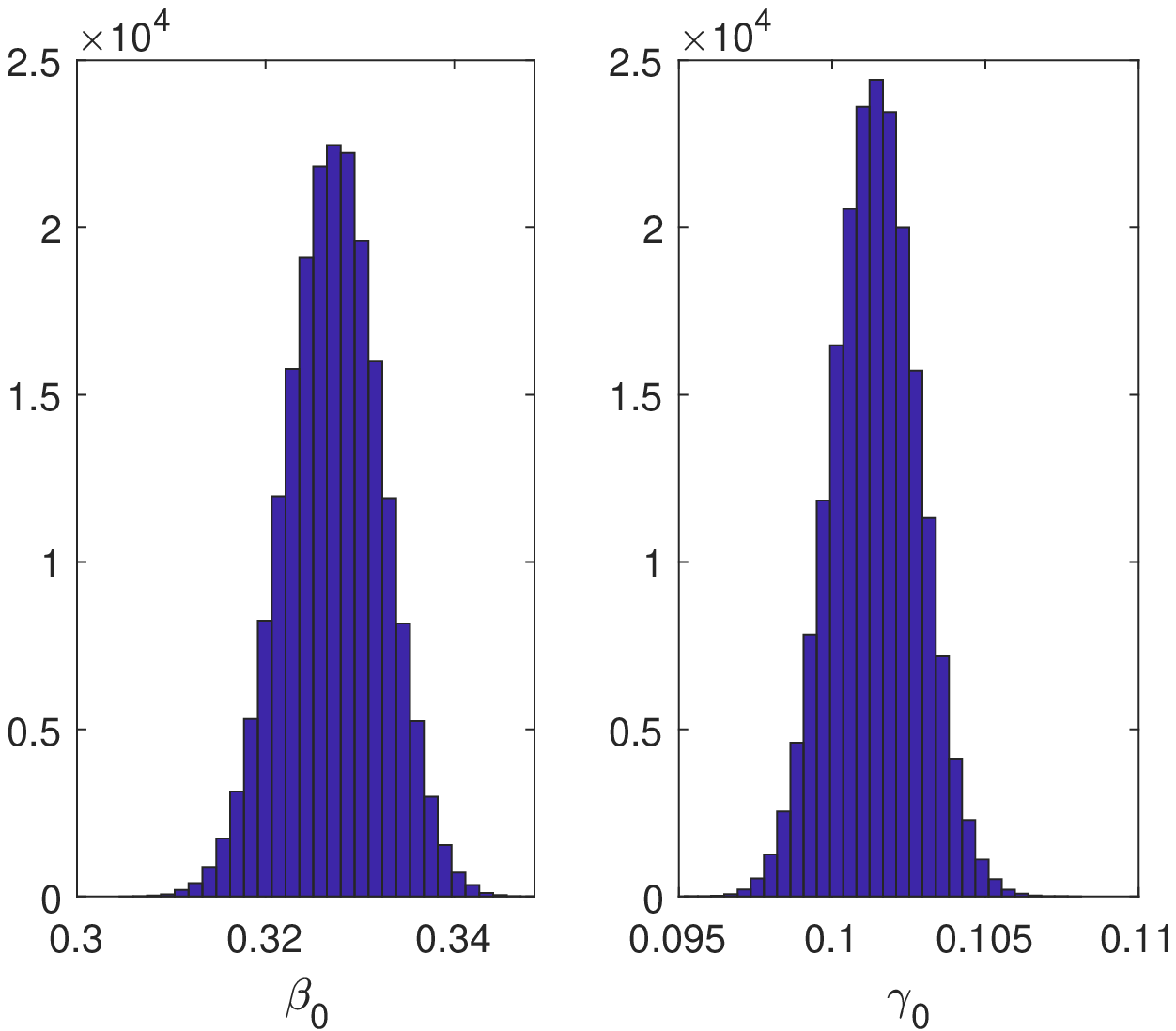}
	\caption{Application of the RBPF algorithm. Left panel: trajectories of infected (red line) and removed (green line). Circles represent the true state. The dynamics up to day 60 are the filtered states, while the dynamics from day 61 to day 67 are forecasts. Right panel: posterior distributions for the parameters $\beta_0$ and $\gamma_0$ at time 60.}
	\label{dinisto_h24}
\end{figure}

The initial guess for the parameter $\theta_0=(\beta_0,\gamma_0)^T$ is $\mu_0=(0.5,0.5)^T$ and the prior covariance matrix $\Sigma_0 = diag(0.05,0.02)$. The mean trajectories of infected and removed people over all the particles obtained running the RBPF algorithm are represented with continuous lines in the left panel of Figure \ref{dinisto_h24} where the circles represent simulated states before the introduction of the observation error. Susceptibles are obtained as $S_t=1-I_t-R_t$ and then the goodness of fit is a consequence of the fit for the other two compartments.

\begin{figure}
	\center
	\includegraphics[scale=0.6]{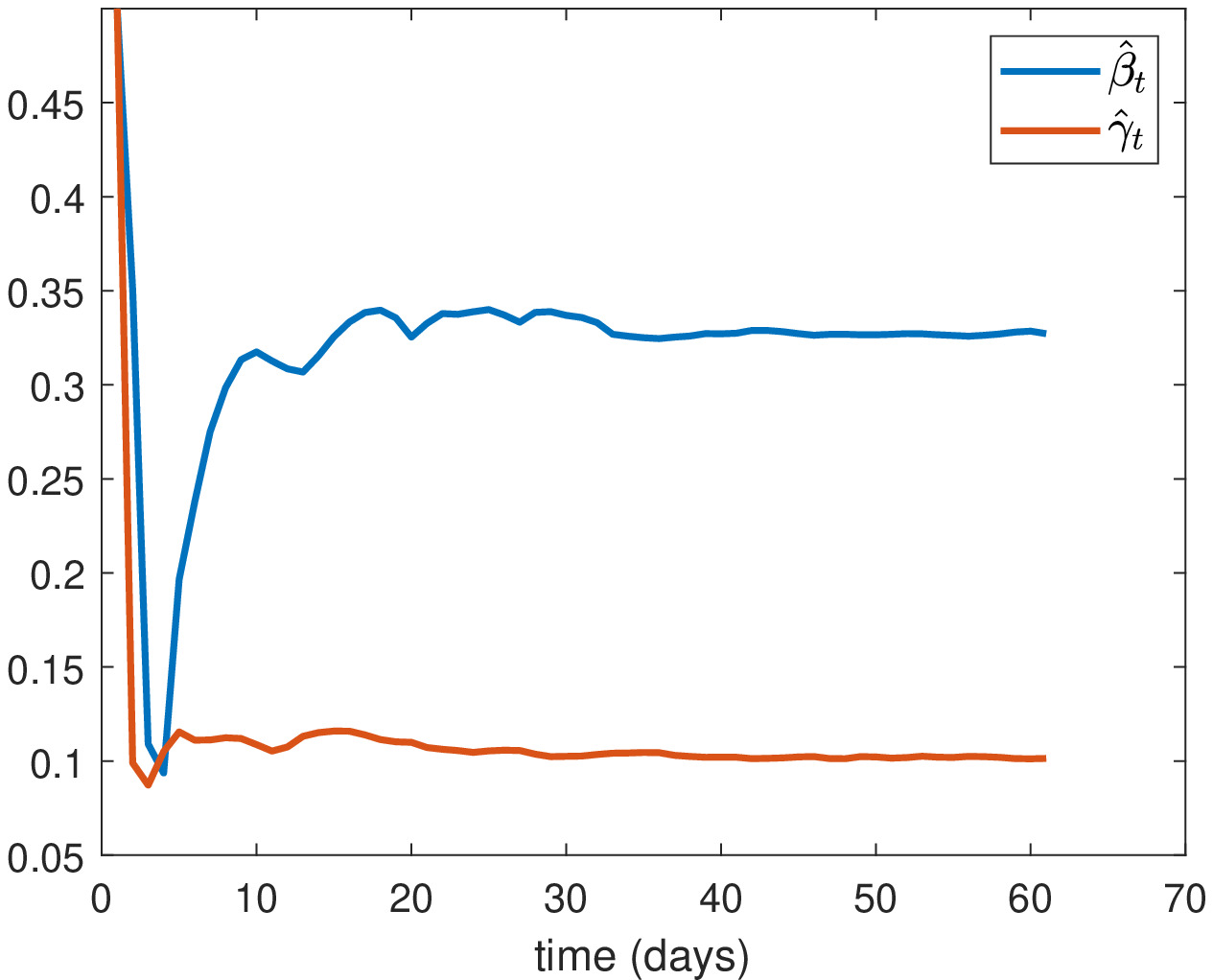}
	\hspace{1cm}
	\includegraphics[scale=0.6]{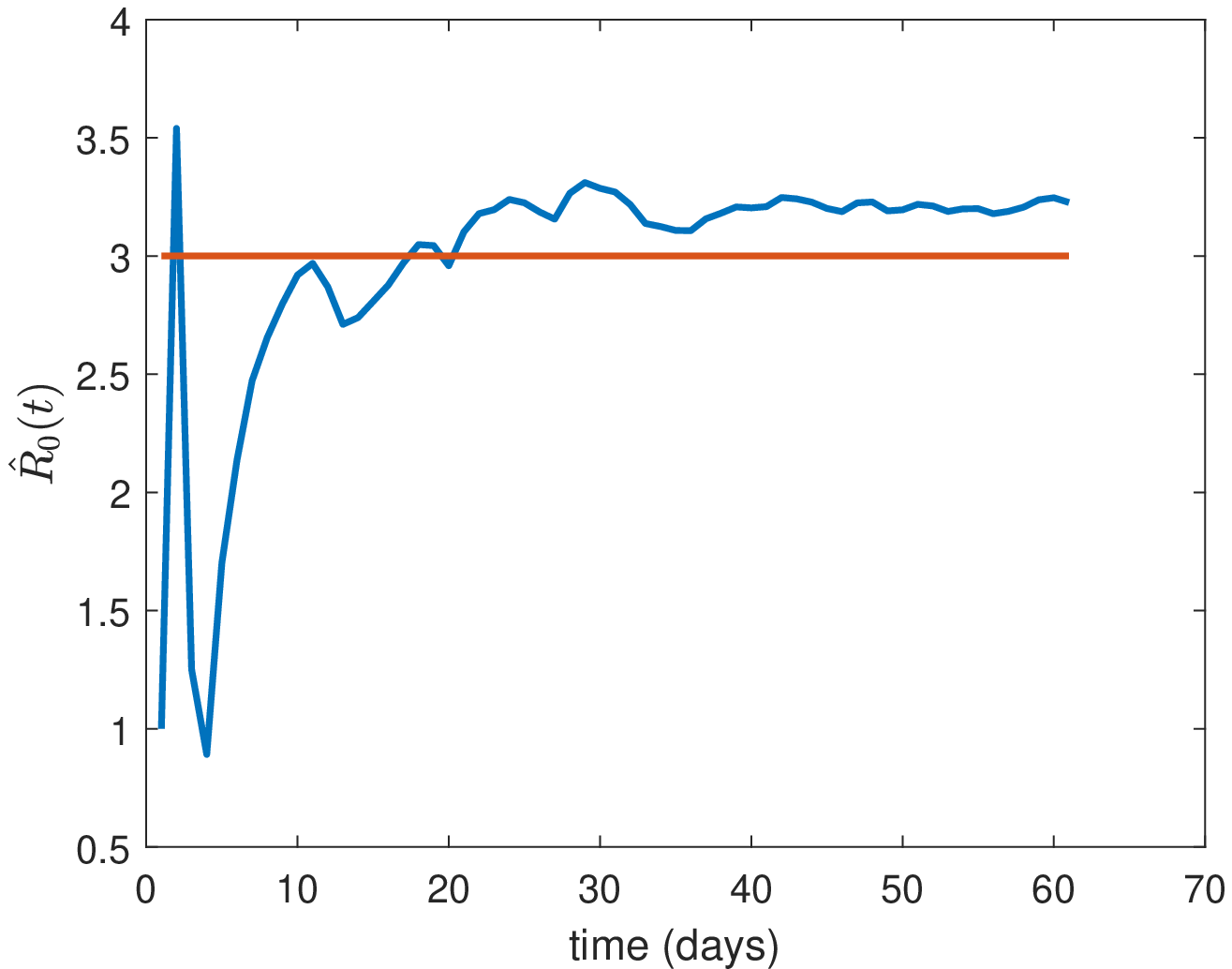}
	\caption{Left panel:  behaviour of $\hat{\beta}_t$ (blue) and $\hat{\gamma}_t$ (red) obtained from \eqref{eq:meantheta}. Right panel: behaviour of $\hat{R}_0(t)$ obtained from \eqref{eq:R0est} (blue line) and value of true $R_0=\beta_0/\gamma_0=0.3/0.1$ (red line).}
	\label{param_h24}
\end{figure}

We denote the estimates of $\beta_0$ and $\gamma_0$ with information up to time $t$ by $\hat{\beta}_t$ and $\hat{\gamma}_t$, see \eqref{eq:meantheta}. Their time behaviour is shown in the left panel of Figure \ref{param_h24}. The behaviour of $\hat{R}_0(t)$, the estimated basic reproduction number \eqref{eq:R0est}, is displayed in the right panel of Figure \ref{param_h24}. 

We denote the filtered or forecasted state as $\hat{I}_t$ and $\hat{R}_t$, where the value of $t$ determines whether we are filtering or forecasting, that is, if our observation period ends at time $s$, then when $t>s$, $\hat{I}_t$ and $\hat{R}_t$ are forecasts; otherwise they are filtered states. From the RBPF we get the filtered states and, using \eqref{eq:forecastMC}, we get a forecast of the dynamics. $\hat{I}_t$ and $\hat{R}_t$ are compared to the true state in the left panel of Figure \ref{dinisto_h24}, where we see that both the fit up to day 60 and the forecast on days 61-67 are satisfactory. In particular, the forecast well represents the trend of the state.

\begin{figure}
\center
\includegraphics[scale=0.8]{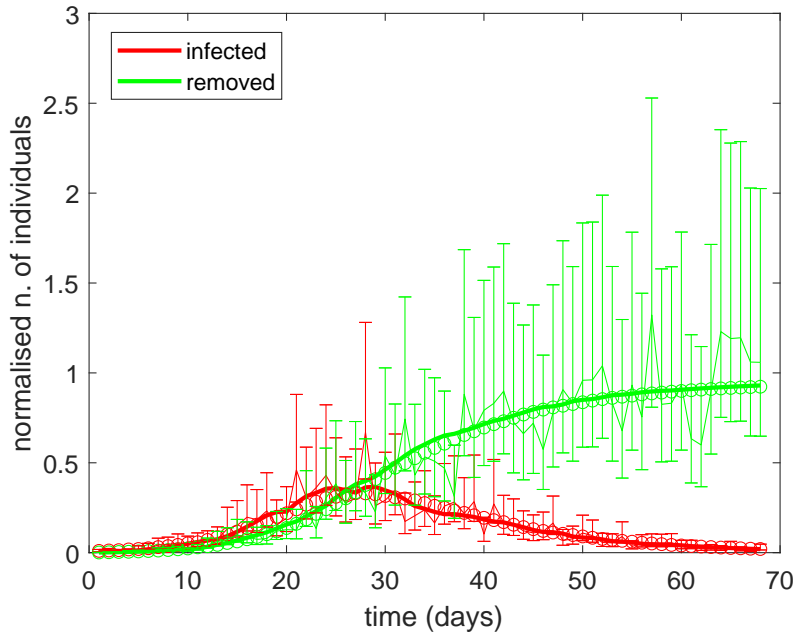}
\caption{True states (circles), filtered state and forecast (thick lines) and adjusted observations (thin line) with 95\% prediction intervals (\ref{eq:cinf}). The thick lines up to day 60 are the filtered states, while those from day 61 to day 67 are forecasts. The thin lines are the observations divided by $u_{0.5}$, the median of the observation error distribution.}
\label{dinamiche_CI}
\end{figure}

In reality the true states are unobserved, and we can only compare the filtered states to the observations, by taking under-detection into account. Therefore, we compute daily prediction intervals for infected $I_t$ and removed $R_t$, as in \eqref{eq:cinf} with $q=0.025$.
In Figure \ref{dinamiche_CI} the prediction intervals are represented by vertical lines, while the thin continuous lines represent the ratio between the observations and the median of the distribution of the observation error (which we may call the adjusted observations).
The width of the prediction intervals reflects the dispersion of the observation error distribution in the left panel of Figure \ref{graf_gamma_dati}, for which $q_{0.025} = 0.10$ and $q_{0.975} = 0.32$. The simulated dynamics (the true states) are inside the intervals and cross the adjusted observations, then, if the adjusted observations and the filtered state agree with each other, this is a necessary condition for the filtered state to follow the true state.

\subsection{Assessment of sample variability}\label{sec:samplevar}
In this section we analyze the variability of the filtered states and of the estimated parameters due to the variability of the data generated from the system (\ref{SIR_stoc_2eq}), in order to get an impression of how far they can get from the true values, even if the true random under-reporting error distribution is used. We use the same parameters of Section~\ref{sec:datafrommodel}, that is, $(\beta_0, \gamma_0) = (0.3, 0.1)$, $\sigma=0.01$, $\eta=0.03$, initial values $\mu_0=(0.5,0.5)^T$ and $\Sigma_0 = diag (0.05,0.02)$ and initial state $X_0=(1\%, 0.1\%)$.

\begin{figure}
\center
\includegraphics[scale=0.55]{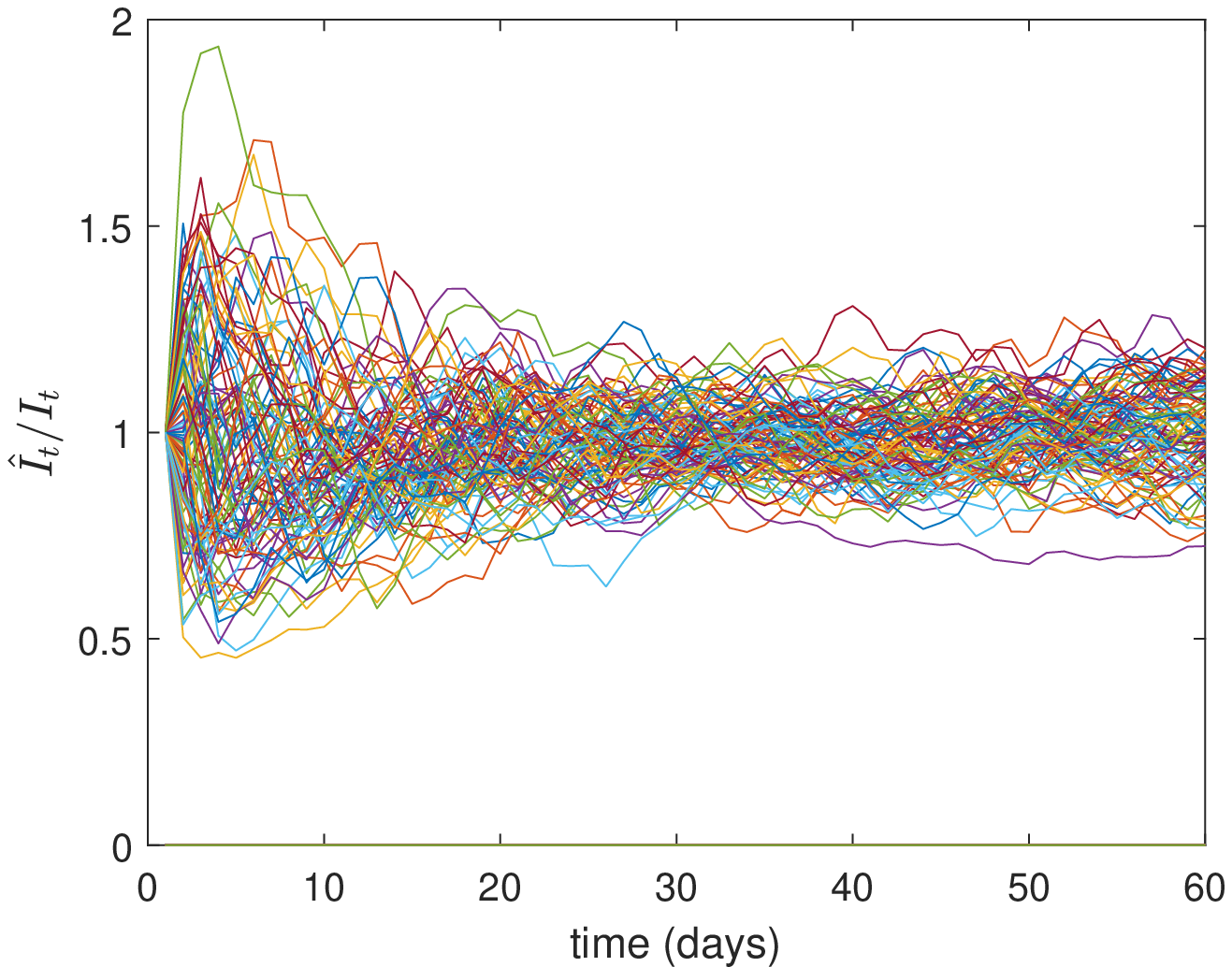}
\hspace{1cm}
\includegraphics[scale=0.55]{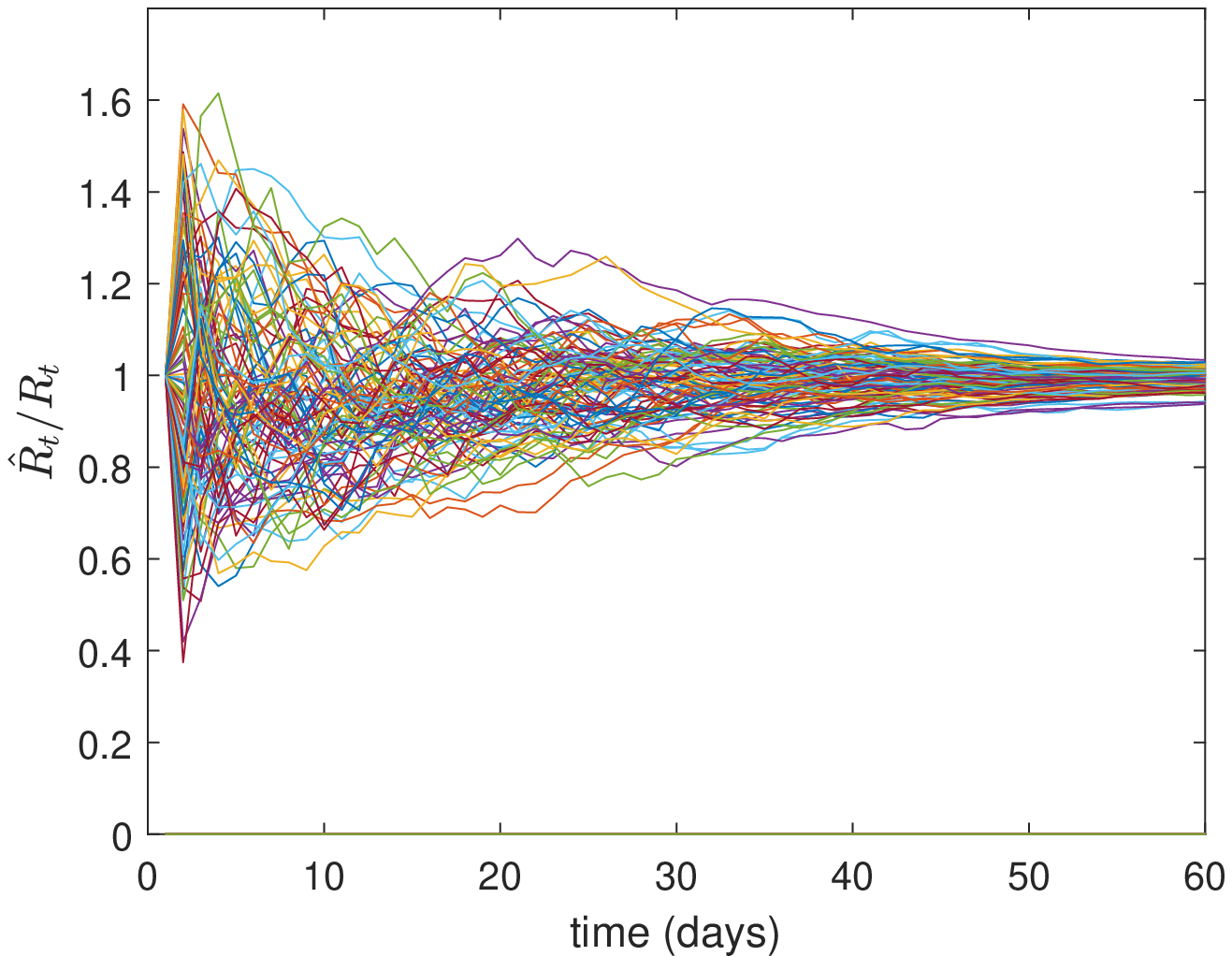}
\caption{Left panel: behaviour of $\hat{I}_t/I_t$ for 100 different simulations. Right panel: behaviour of $\hat{R}_t/R_t$ for 100 different simulations.}
\label{rapporti}
\end{figure}

We generate 500 dynamics from the system (\ref{SIR_stoc_2eq}), from which we obtain 500 trajectories of observed infected and removed people. Then, we run the RBPF algorithm on every simulated dataset, with a time step of $1/24$  day and for 200,000 particles.
For every simulations we compute the trajectories $\hat{I}_t/I_t$  and $\hat{R}_t/R_t$, where we recall that $\hat{I}_t$ and $\hat{R}_t$ are the estimated infected and removed individuals filtered by the RBPF algorithm, while  $I_t$ and $R_t$ are the true states for the corresponding simulation. For the sake of a clear representation, in Figure \ref{rapporti} we show one every five trajectories of $\hat{I}_t/I_t$ (left panel) and  $\hat{R}_t/R_t$ (right panel).

\begin{figure}
\center
\includegraphics[scale=0.6]{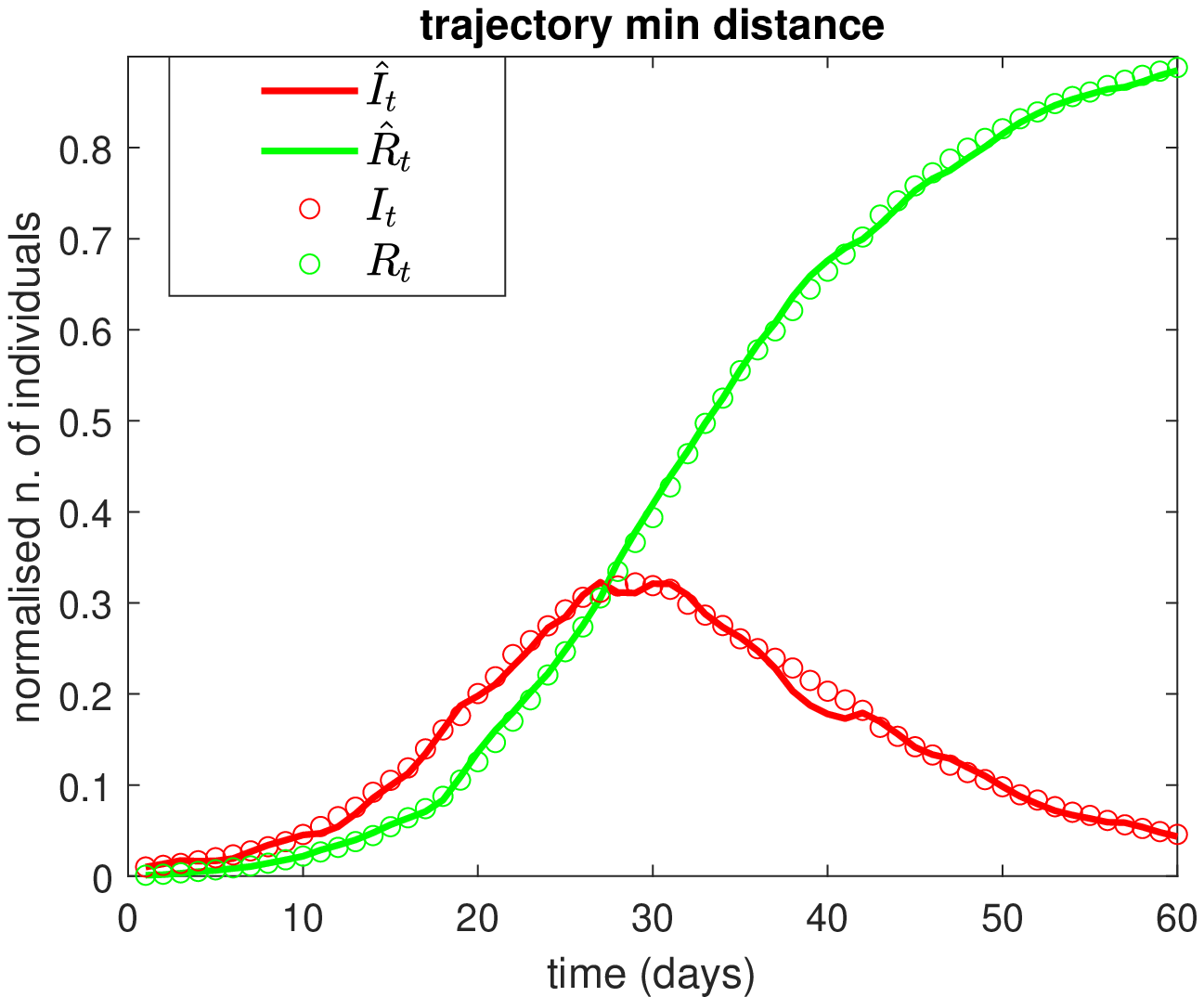}
\hspace{1cm}
\includegraphics[scale=0.6]{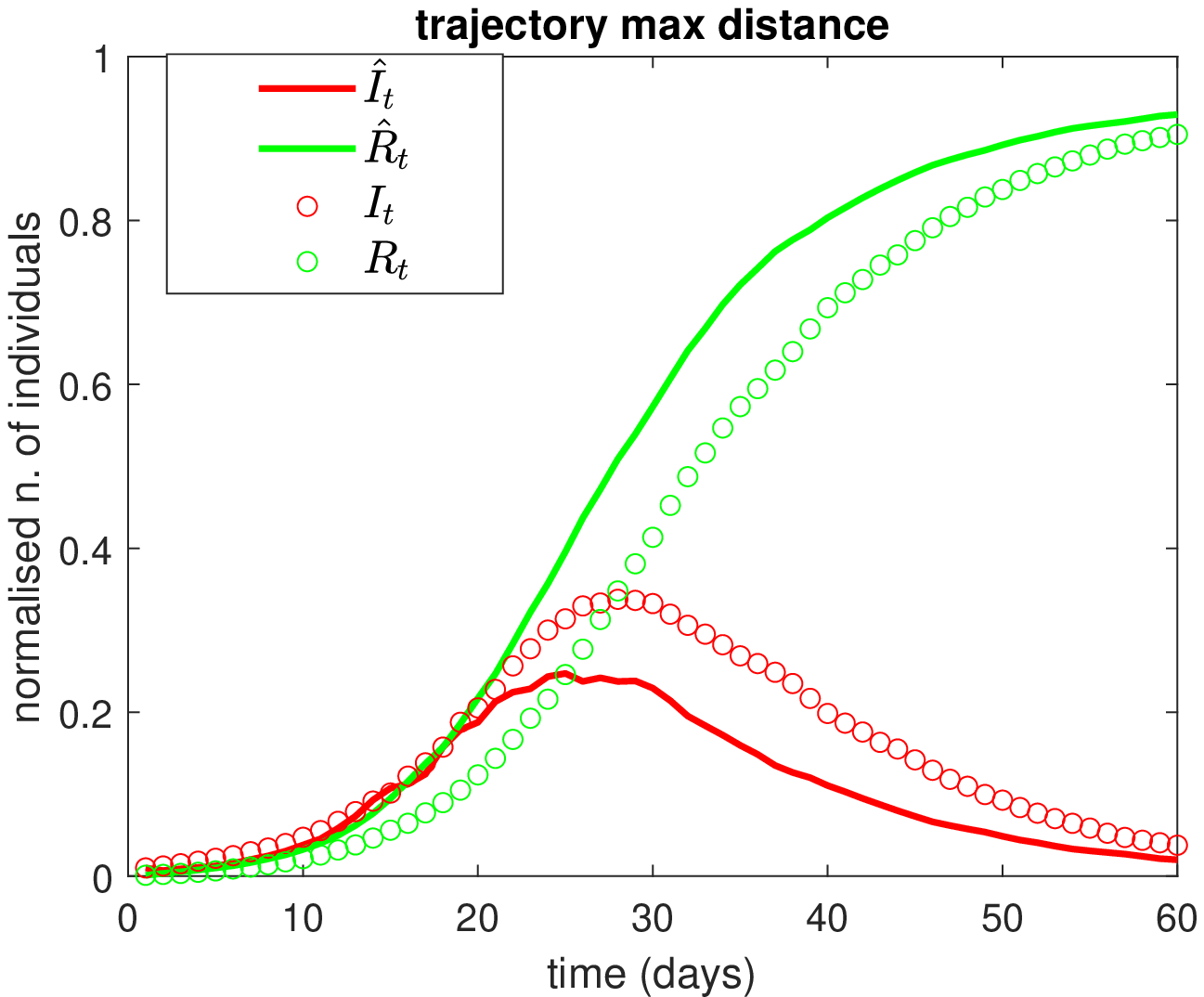}
\caption{Left panel: dynamics of filtered states $\hat{I}_t$ and $\hat{R}_t$ (continuous lines) in the case of minimum sum of root mean square error between filtered states and true states (circles). Right panel: dynamics of filtered states $\hat{I}_t$ and $\hat{R}_t$ (continuous lines) in the case of maximum sum of root mean square error between filtered states and true states (circles).}
\label{traiettminmax}
\end{figure}

In both panels of Figure \ref{rapporti}, after a transient phase with larger dispersion, $\hat{I}_t/I_t$ and $\hat{R}_t/R_t$ end up fluctuating around 1, with a stable dispersion in the left panel and a decreasing dispersion in the right panel. 
Now consider the sum of the root mean square error (RMSE) between $\hat{I}_t$ and $I_t$ and of the RMSE between $\hat{R}_t$ and $R_t$ for all the trajectories, as a measure of distance between the estimate and the truth. Among all the trajectories we represented the one with the smallest and the one with the largest distance in the left and in the right panels of Figure \ref{traiettminmax}, respectively. The latter picture shows that the fit may be very unsatisfactory. 

\begin{figure}
\center
\includegraphics[scale=0.8]{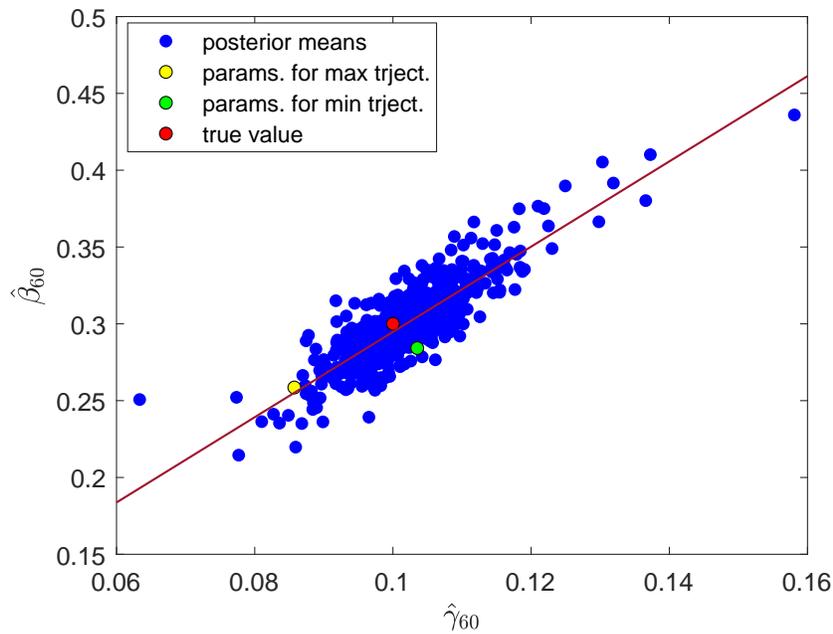}
\caption{Scatter plot of the parameters $(\hat{\gamma}_{60},\hat{\beta}_{60})$ obtained for the different simulations. The red dot represents the true pair $(0.1,0.3)$. The green dot represents the pair of parameters relative to the filtered state trajectory with minimum distance from the true states (left panel of Figure \ref{traiettminmax}). The yellow dot represents the pair of parameters corresponding to the filtered state trajectory with maximum distance from the true states (right panel of Figure \ref{traiettminmax}). The line is the least squares fit of $\hat{\beta}_{60}$ against $\hat{\gamma}_{60}$  (slope 2.78).}
\label{scatterpar}
\end{figure}

Finally, we report the scatter plot of all the pairs $(\hat{\beta}_t,\hat{\gamma}_t)$ obtained in the 500 simulations (Figure \ref{scatterpar}) at $t=60$. We observe that they are dispersed around the true value of the parameter $(0.3,0.1)$. The pair corresponding to the trajectory of minimum distance (green dot) is closer to the true parameter (red dot) than the pair estimated from the trajectory of maximum distance (yellow dot). Then a good fit to the state trajectory is associated with a better estimate of unknown parameters, provided the remaining parameters, which have been assumed as known, are correctly assigned, as we will see in the next section. 

An interesting feature of Figure \ref{scatterpar} is that the ratio $\hat{\beta}_t/\hat{\gamma}_t$ shows a smaller variability than $\hat{\beta}_t$ and $\hat{\gamma}_t$, around a straight line with slope close to three, the true value of $R_0$. Therefore we expect to be able to estimate the basic reproduction number with better accuracy and precision than the infection and removal rate parameters.

\subsection{Identifiability}\label{sec:ident}
A statistical model, belonging to a parametric family, is called identifiable if, for any two different values of parameters, there exists at least a measurable set in the sample space that is not assigned the same probability by the two members of the family, that is, given $\theta_{01}\ne\theta_{02}$, there exists at least one set $B$ such that
\begin{equation}\label{eq:identifiability}
\Pr(Y_{1:t}\in B;\theta_{01})\ne \Pr(Y_{1:t}\textbf{}\in B;\theta_{02}) \ ,
\end{equation}
where $Y_{1:t}$ denotes a finite-length trajectory of observations from \eqref{errore_oss}. 
For a deterministic model, this property is called structural identifiability, which holds if there exists a map from the parameter to the output $\theta_0 \mapsto y_{1:t}(\theta_0)$ which is injective, that is, given $\theta_{01}\ne\theta_{02}$, the two models $y(\theta_{01})$ and $y(\theta_{02})$ describe different output trajectories. By a differential algebra approach, \cite{piazzola2020note} show that the following deterministic SIR model with its output
\begin{equation}
	\label{SIRdet_tamell}
	\left\{ \begin{array}{l}
		\frac{dI_t}{dt}=\beta_0 I_t (N-I_t-R_t) - \gamma_0 I_t\\
		\frac{dR_t}{dt}=\gamma_0 I_t\\
		y_{1,t} = \frac{1}{K} I_t\\
		y_{2,t} = \frac{1}{K} R_t\\
	\end{array}\right.
\end{equation}
defined for a non-normalised population of size $N$, is structurally identifiable with respect to the unknown parameters $\beta_0$, $\gamma_0$ and $K$. The parameter $K>1$ accounts for under-reporting of infected and recovered and has the same function of the $U$ random variables in \eqref{errore_oss}. Then, after adding noise to the output they go on to show that, despite structural identifiability, the parameters may not be practically identifiable, that is, a good or acceptable agreement between observations and fit is displayed for different values of the parameters when observations end before reaching the peak.

\begin{figure}
\center
\includegraphics[scale=0.6]{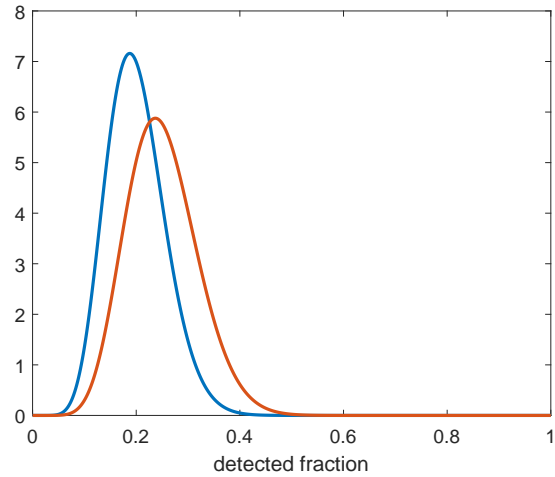}
\caption{Comparison between the two beta densities used to model the observation error.}
\label{confr_beta}
\end{figure}

The way randomness has been included into this problem via model \eqref{SIR_disc}-\eqref{errore_oss} is different from \cite{piazzola2020note}, but we also observe practical identifiability. The problem of identifiability is also discussed in \cite{ganyani2021} for stochastic SIR models.
We generate state trajectories composed by 30 daily values using model (\ref{SIR_stoc_2eq}) with parameters $\beta_0=0.1$ and $\gamma_0=0.03$. Then, to obtain the actual observations we multiply each value by a number drawn from a beta distribution with parameters $a=10$ and $b=40$ (blue line in Figure \ref{confr_beta}). After running the RBPF algorithm using the same beta distribution we obtain results analogous to those in the previous sections, that is, a satisfactory fit of the observed dynamics and a good estimate of the parameters $\beta_0$ and $\gamma_0$.

\begin{figure}
\center
\includegraphics[scale=0.55]{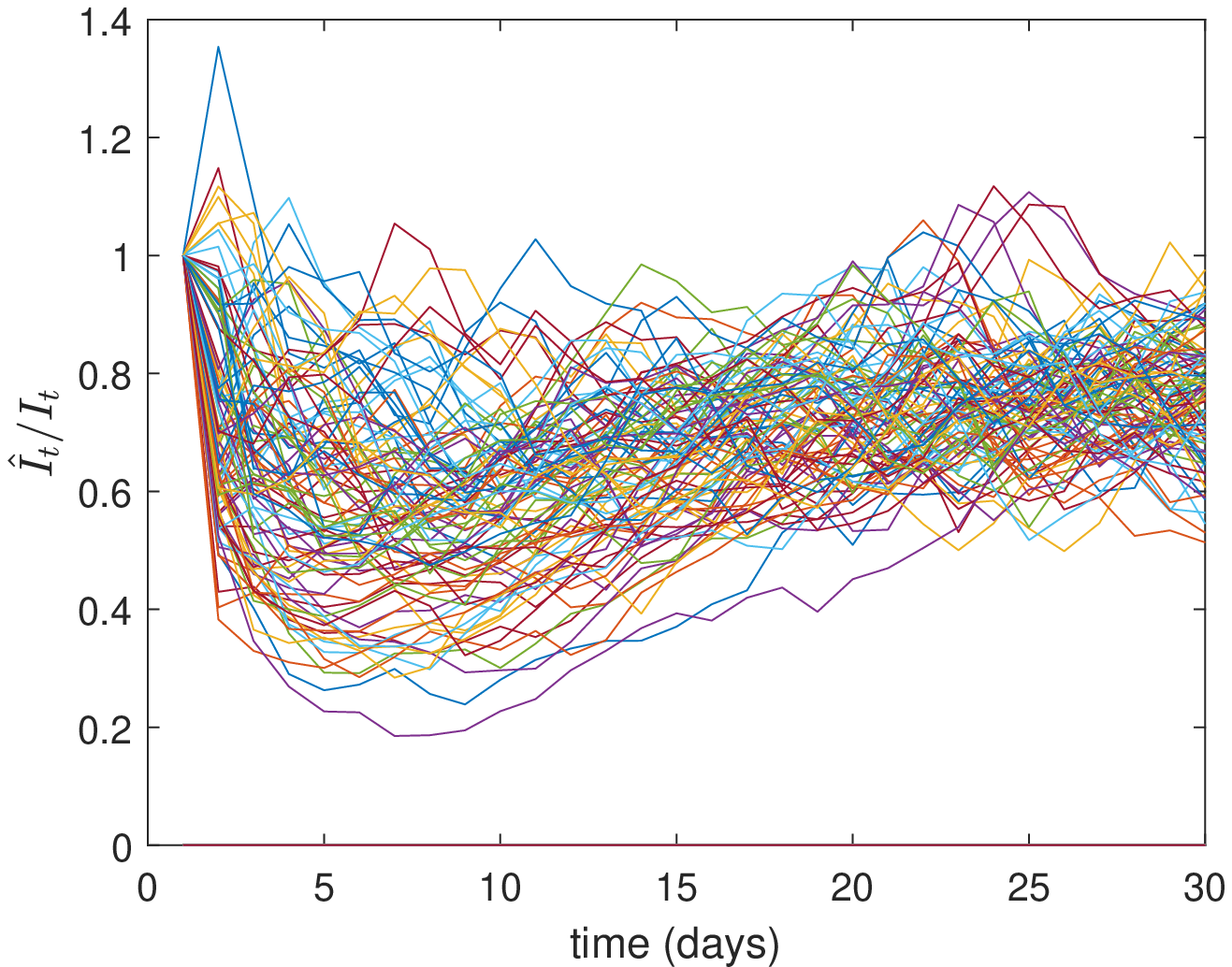}
\hspace{1cm}
\includegraphics[scale=0.55]{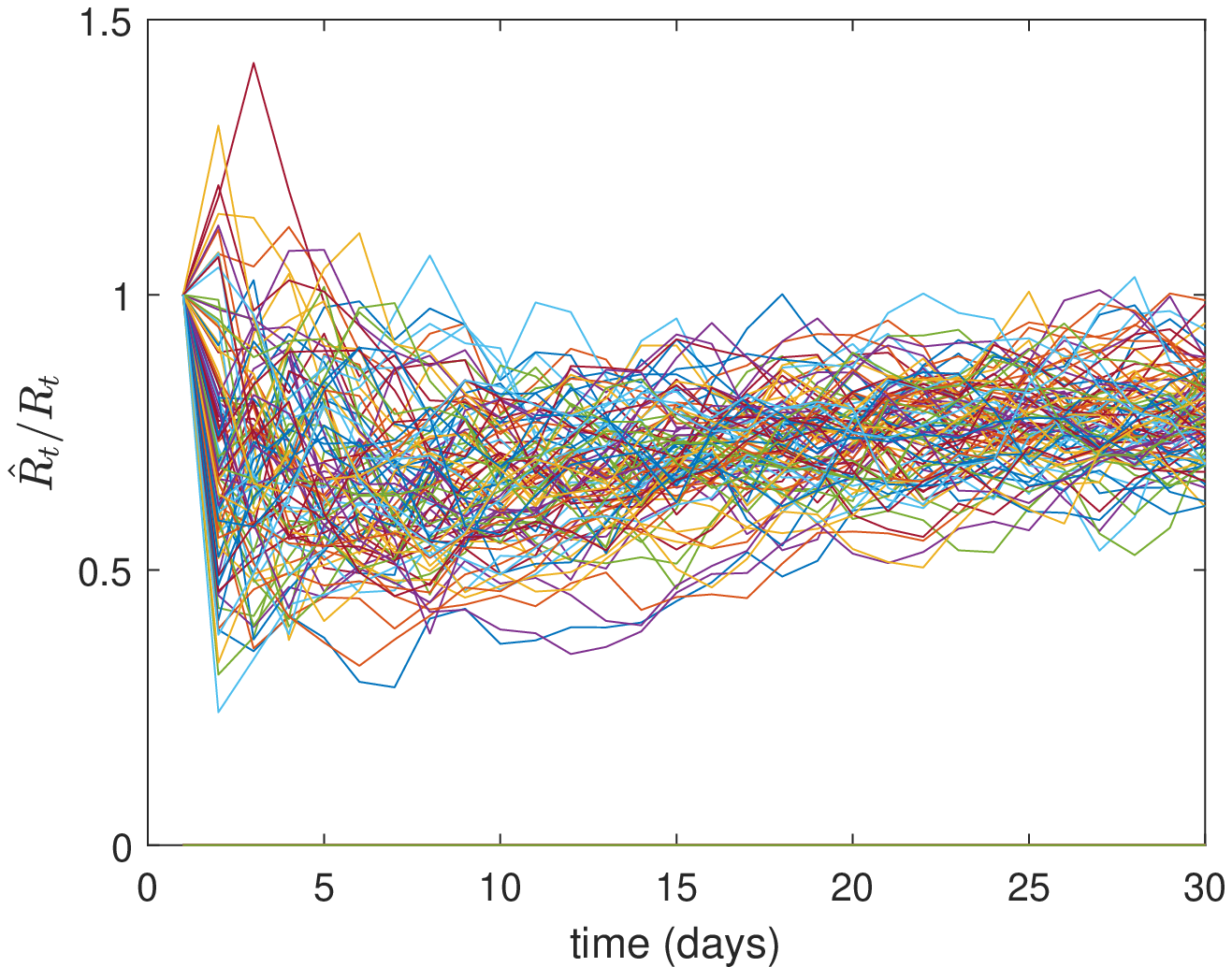}
\caption{Left panel: behaviour of $\hat{I}_t/I_t$ for 100 different simulations. Right panel: behaviour of $\hat{R}_t/R_t$ for 100 different simulations.}
\label{rapportiIR_media06}
\end{figure}

Now we run the RBPF algorithm assuming an observation error with beta distribution with a mean larger than the truth, with parameters $a=10$ and $b=30$ (red line in Figure \ref{confr_beta}). We compare the filtered states with both the true ones and the observed data.  First, we consider 500 simulations and look at the ratio between the filtered state and the state. For the sake of a clear representation, in Figure \ref{rapportiIR_media06} we show one every five trajectories for both the infected and the removed individuals. These ratios are, generally, less than 1 denoting an underestimation of both infected and removed.

\begin{figure}
\center
\includegraphics[scale=0.55]{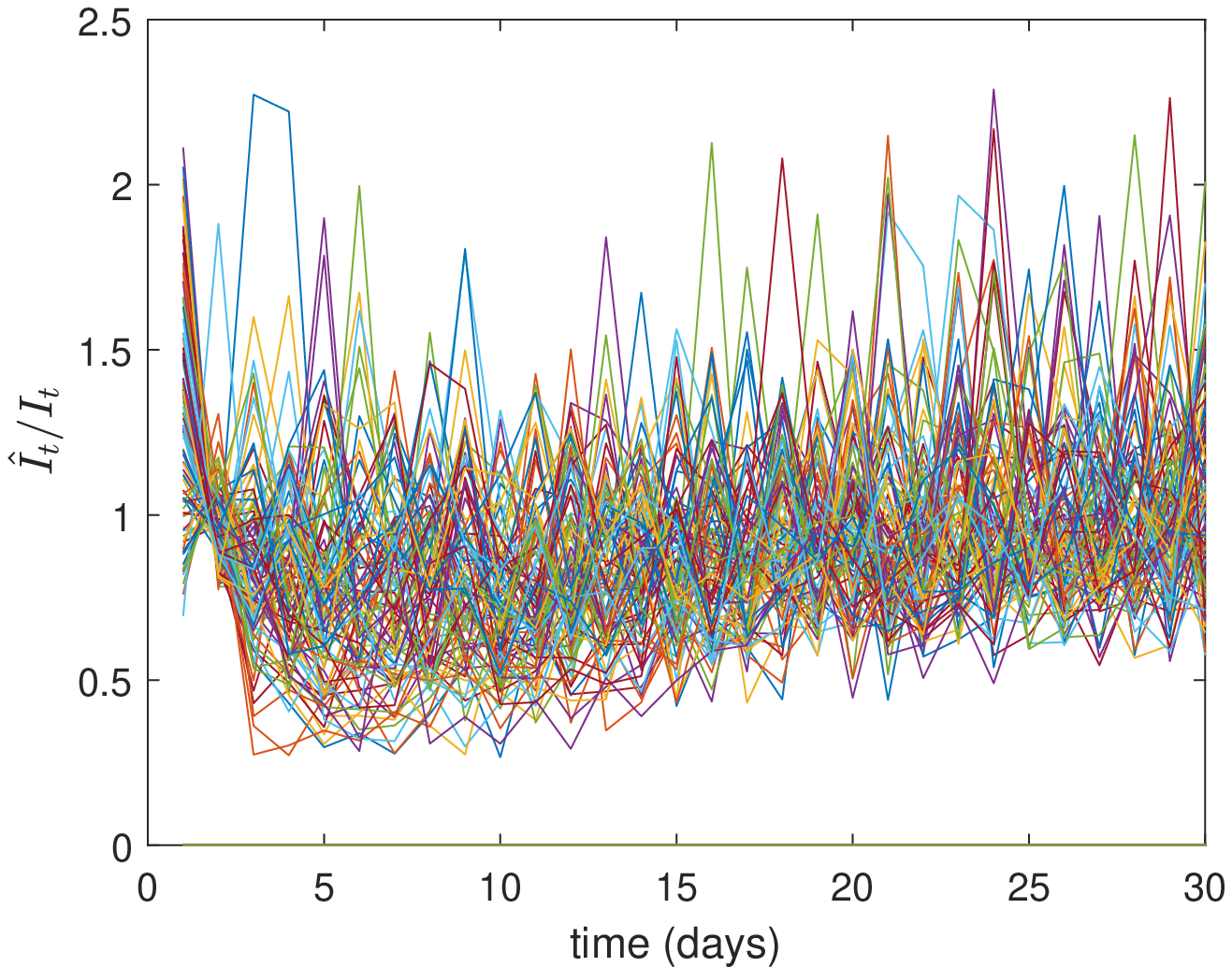}
\hspace{1cm}
\includegraphics[scale=0.55]{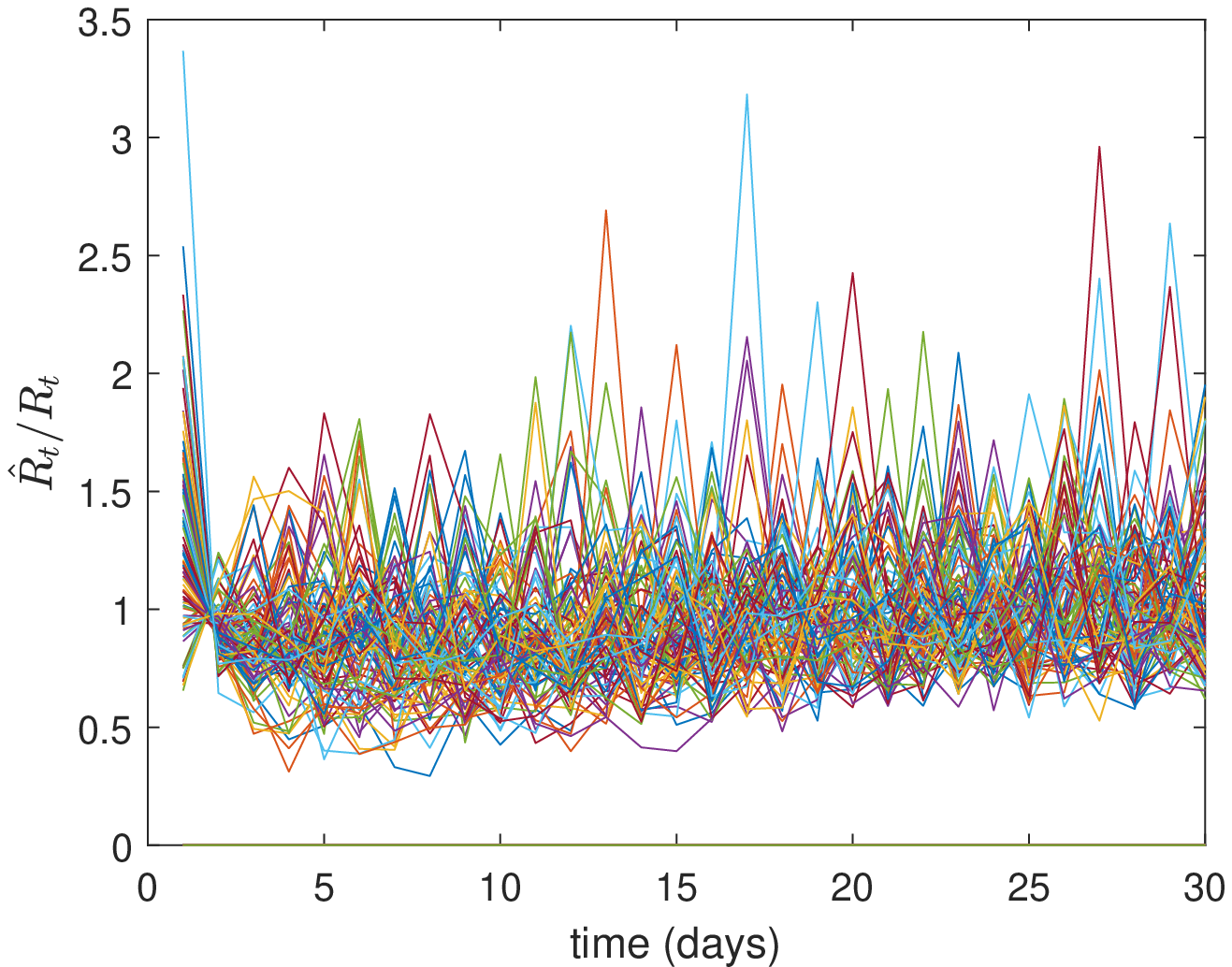}
\caption{Left panel: behaviour of $\hat{I}_t/(y_{t,1}/m)$ for 100 different simulations, where $m$ is the true median of the observation error distribution. Right panel: behaviour of $\hat{I}_t/(y_{t,2}/m)$ for 100 different simulations.}
\label{rapportiIRobs_media06}
\end{figure}

\begin{figure}
\center
\includegraphics[scale=0.55]{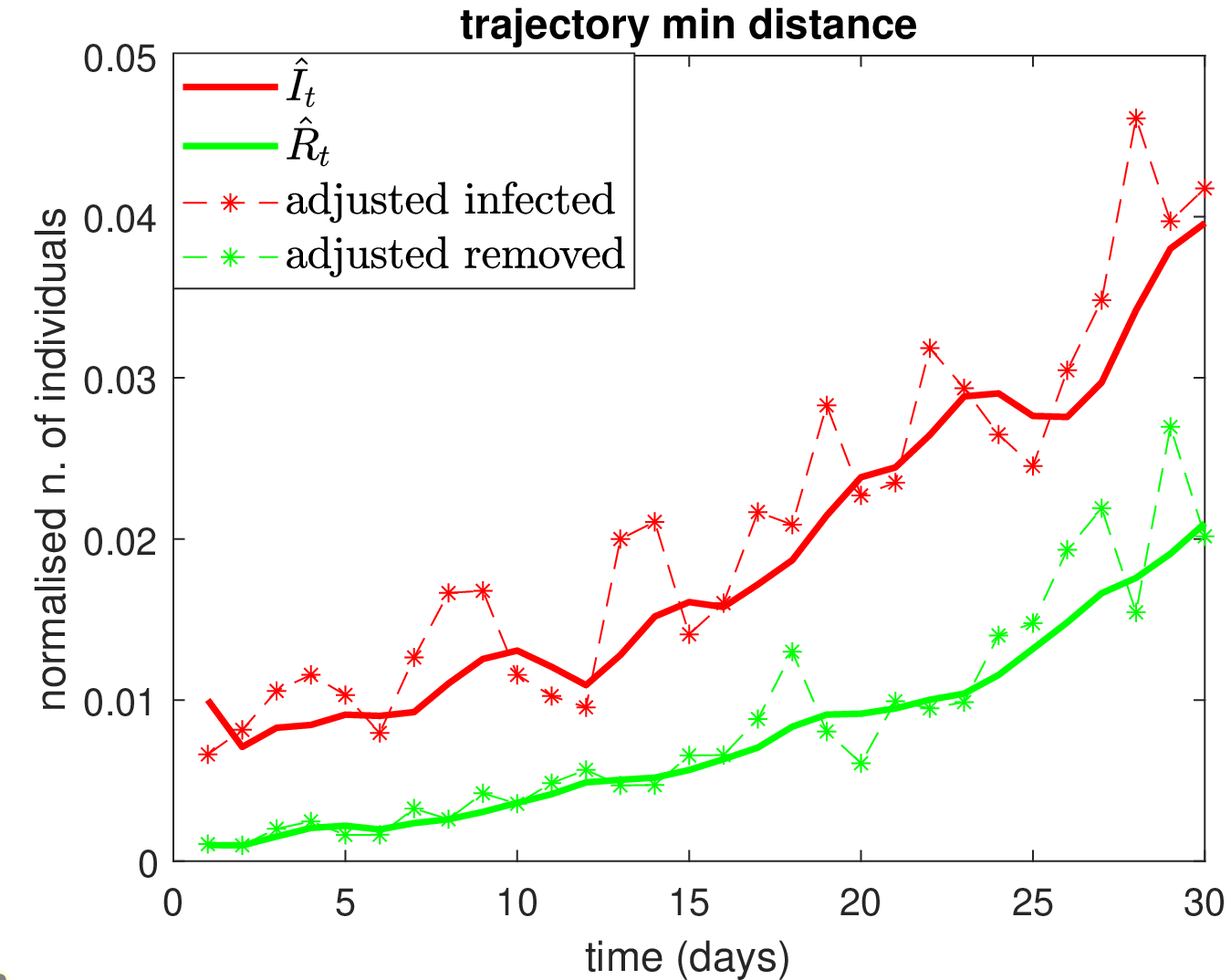}
\hspace{1cm}
\includegraphics[scale=0.55]{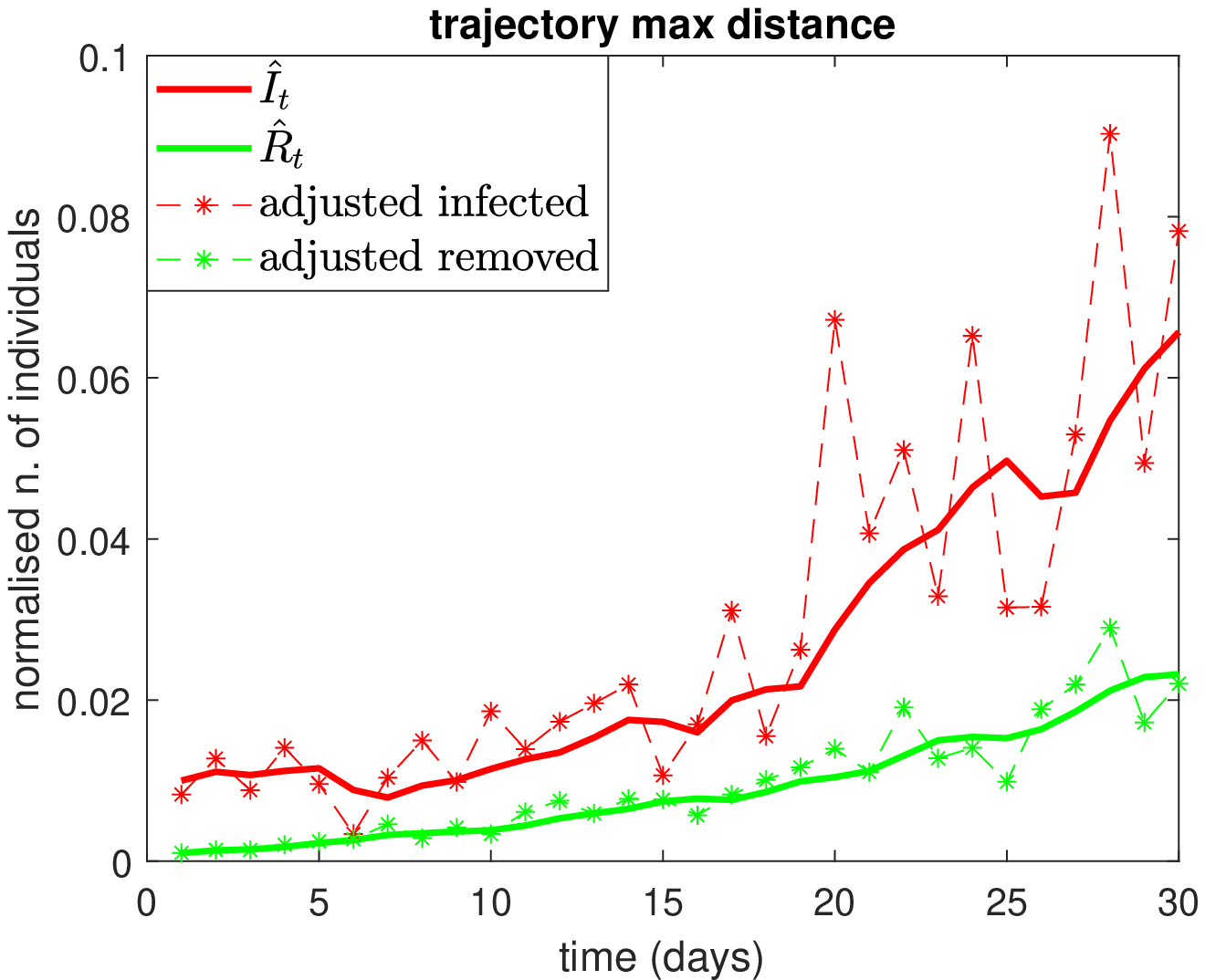}
\caption{Left panel: case of minimum sum of root mean square error between true and filtered state, dynamics of $\hat{I}_t$ and $\hat{R}_t$ (continuous lines) and adjusted observations (thin line with asterisks). Right panel: case of maximum sum of root mean square error between true and filtered state, dynamics of $\hat{I}_t$ and $\hat{R}_t$ (continuous lines) and adjusted observations (thin line with asterisks).}
\label{dinam_min_media06}
\end{figure}

Then, we consider the ratio between the filtered states and the adjusted observations. The results of one every five trajectories are reported in Figure \ref{rapportiIRobs_media06} for both infected and removed individuals. These ratios fluctuate around 1, denoting a satisfactory fit to the observed data. Figure \ref{dinam_min_media06}  shows the dynamics of two simulations: the dynamics with minimum distance of the filtered state from the true state (intended as the minimum sum of the root mean square errors over the two components), in the left panel, and the dynamics with maximum distance, in the right panel. Both trajectories fit the (scaled) observed data reasonably well. 

\begin{figure}
\center
\includegraphics[scale=0.8]{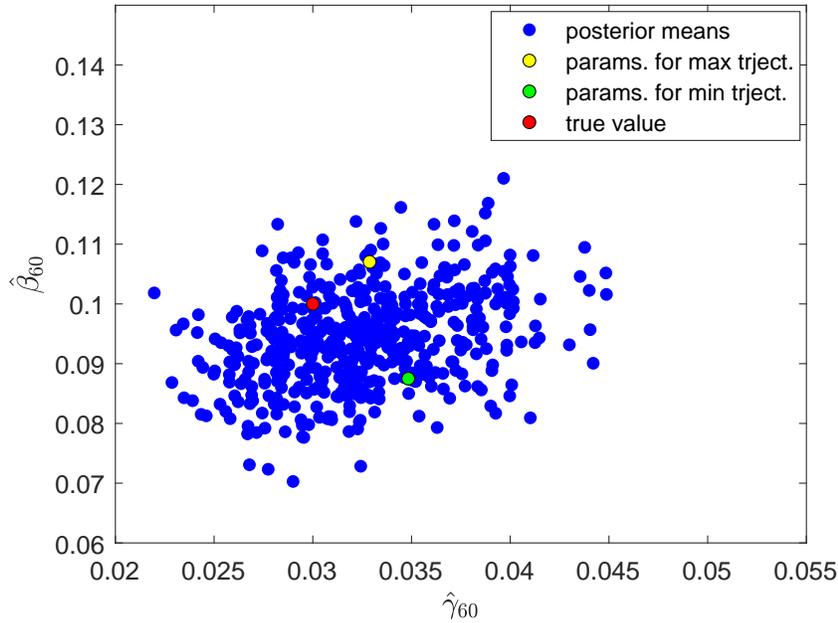}
\caption{Scatter plot of the parameters $(\hat{\gamma}_{60},\hat{\beta}_{60})$, obtained for the different simulations, with wrong parameters $a$ and $b$ in the observation error distribution. The red dot represents the true pair $(0.03,0.1)$. The green dot is the estimate corresponding to minimum distance case (left panel of Figure \ref{dinam_min_media06}). The yellow dot is the case corresponding to the maximum distance case (right panel of Figure \ref{dinam_min_media06}).}
\label{scatter_param_media06}
\end{figure}

Even if the filtered states follow the adjusted observations, the values of the estimated parameters are not correct, as shown in Figure \ref{scatter_param_media06}. In fact, the pair of estimated parameters in the 100 simulations are not equally dispersed around the true value (red point) but are placed mainly below the true value, denoting a bad estimation for $\beta_0$. The estimation of $\gamma_0$ is better.

It follows that it is very important to suitably choose the beta distribution of the observation error (as we have done in Section \ref{sec:realdata}) in the collection of infected and removed people to avoid practical nonidentifiability.

\section{Real data}\label{sec:realdata}
\begin{figure}
\begin{center}
\includegraphics[scale=0.8]{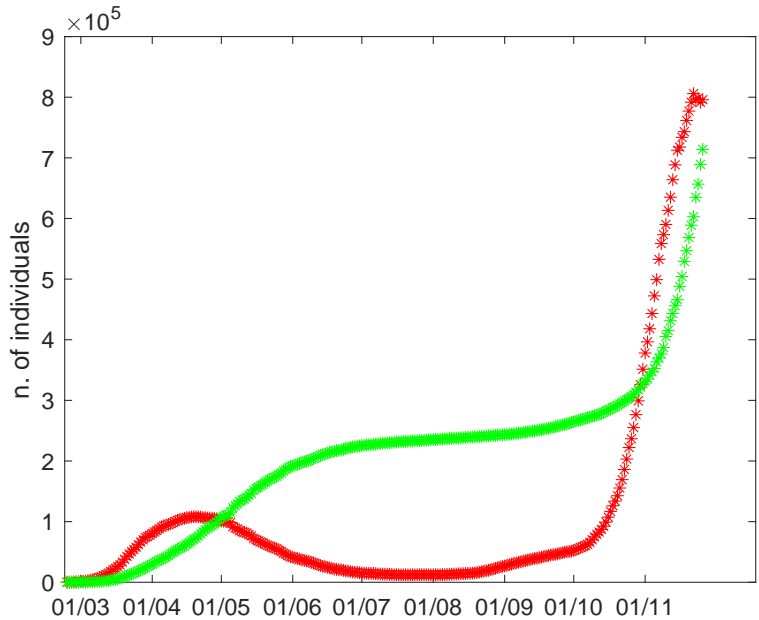}
\caption{Infected (red asterisks) and removed (green asterisks) in Italy from 24th February (time 0) to 26th November 2020. Data from Protezione Civile.}
\label{fig_infetti_rimossi}
\end{center}
\end{figure}

In this section we use data of the epidemic in Italy, collected by Protezione Civile (Civil Protection Department) from 24th February. As a first step we consider the data for the whole Italy from 1st March up to 26th November 2020. Available data are the number of infected, dead, and recovered individuals. Removed people can be obtained by summing dead and recovered people. In Italy, all deaths of people infected with SARS-CoV-2 were classified as COVID-19 (\cite{riccardo2020epidemiological}). The infected and removed individuals in Italy from  24th February to 26th November are represented in Figure \ref{fig_infetti_rimossi}. The total residing population as of 31st December 2019 is 60,244,639 people, as certified by Istat. 

To deal with underdetection we consider observations as generated by (\ref{errore_oss}), where we recall that $U_{t,1}$ and $U_{t,2}$ are independently beta distributed with common shape parameters $a$ and $b$.
In particular, we have $I_{obs}=Y_1= U_1 X_1$ and $R_{obs}= Y_2 = U_2 X_2$, where the additional notations $I_{obs}$ and $R_{obs}$ are introduced as meaningful names for what follows.

As we can see from Figure \ref{fig_infetti_rimossi} the epidemic can be divided in two waves: the first officially began on 24th February and lasted until mid-summer, when the number of infected people began to rise again, as in the rest of Europe (\cite{bontempi2020europe}). The second wave 
is distinguished from the first also by the increased test capacity. Hence we consider the two waves as different models, with respect to both the SIR parameters and the observation error distribution parameters and we conventionally set the start of the second wave on 1st August.

\subsection{Assigning parameters of the observation error distribution}\label{sec:assegnaerrdist}
To fix $a$ and $b$ we refer to the Infection Fatality Ratio (IFR), a fundamental quantity to estimate the 
seriousness of the epidemic, and to its crude estimate, 
the Case Fatality Ratio (CFR). The IFR is the ratio of COVID-19 deaths to total infections of SARS-CoV-2, 
asymptomatic and undiagnosed infections included, while the CFR is the ratio of COVID-19 deaths to confirmed cases. 
By the definition, the CFR is greater than the IFR. At any time $t$ an estimate $CFR_t$ of the CFR and an estimate $IFR_t$ of the IFR are related by the simple relationship
\begin{equation}\label{eq:cfrifr}
CFR_t = \frac{D_t}{R_{obs,t}}=\frac{D_t}{R_t}\times \frac{R_t}{R_{obs,t}}=IFR_t\times \frac{R_t}{R_{obs,t}}
\end{equation}
where $D_t$ denotes all deaths by time $t$ (\cite{ghani2005methods}).

Since by \eqref{eq:cfrifr}
\begin{equation}\label{eq:robsrt}
	R_{obs,t} = \frac{IFR_t}{CFR_t} \times R_t \ ,
\end{equation}
the ratio $IFR_t/CFR_t$ can be regarded as the under-reporting factor that we modelled as the $U$ beta random variable introduced earlier.	

Given estimates of $IFR_t$ as $t=1, \ldots, T$ and the corresponding observed sequence of $CFR_t$ we would obtain a sample $u_1=IFR_1/CFR_1,\ldots,u_T=IFR_T/CFR_T$ and an estimate of $a$ and $b$ by any established method. Using the method of moments, for example, and considering an estimate of the IFR to substitute $IFR_t$, we would get
\begin{equation}\label{eq:mm_rlambda}
	\begin{aligned}
		\hat{a} & = \bar{u}\left(\frac{\bar{u}(1-\bar{u})}{s^2_u}-1\right) 
		= \bar{u}_{1/CFR}\left\{\frac{\bar{u}_{1/CFR}(1-IFR\, \bar{u}_{1/CFR})}{s^2_{1/CFR}}-IFR\right\} \\
		\hat{b} & = (1-\bar{u})\left(\frac{\bar{u}(1-\bar{u})}{s^2_u}-1\right) 
		= \frac{1-IFR\,\bar{u}_{1/CFR}}{IFR}\left\{\frac{\bar{u}_{1/CFR}
		(1-IFR\, \bar{u}_{1/CFR})}{s^2_{1/CFR}}-IFR\right\}
	\end{aligned}
\end{equation}
where $\bar{u}$ and $s^2_u$ are the sample mean and variance of $(u_1,\ldots,u_T)$ and $\bar{u}_{1/CFR}$ and $s^2_{1/CFR}$ are the sample mean and variance of $1/CFR_1,\ldots,1/CFR_T$. These equations show that the IFR affects both parameters.

The fatality ratio approach has the advantage that the IFR is a pure number and information on its value can be gathered from different populations. Then, in practice, we may estimate the sample mean and variance of $1/CFR_1,\ldots,1/CFR_T$ from the observed fatality and removal data, and for a selected $IFR$ assign $\hat{a}$ and $\hat{b}$. If a range of values is available for the IFR from another source, such as a confidence or a credibility interval, we may repeat the analysis for $IFR$ varying within the interval and evaluate the sensitivity of the results.

For Italy, we may use an indirect method to point at a plausible value of the IFR within this interval, taking advantage of a seroprevalence survey targeting IgG antibody conducted in Italy from May to July by Istat, the Italian national statistical office, and the Italian Health Ministry. Preliminary results obtained from 64,660 people were presented in early August (\cite{istat2020agg}). According to them, almost 1.5 million people in Italy or 2.5$\%$ of the population had developed coronavirus antibodies, a figure six times larger than official numbers reported. In short, the idea is to compare the 2.5$\%$ figure of people who developed antibodies to the healed people (who have antibodies) estimated from the filtered state $\hat{R}_t$ in an appropriate time interval. The infected compartment may also contain seropositive individuals, still, the fraction of people in this compartment had become small when Istat's survey started, so we consider only the recovered compartment. The reasoning behind this comparison is that if the assumed IFR is correct, then the observation error distribution derived from \eqref{eq:mm_rlambda} is correct and the filtered states are realistic and they should be in agreement with the Istat survey result.

To be more specific, let $R_t = H_t + D_t$, where $H_t$ and $D_t$ are the fractions (over the population) of healed and dead people by time $t$, respectively. Healed people can be seronegative if IgG antibodies are no longer in their system, but we can safely assume that a person enters the healed record soon so they s/he can be considered as seropositive when they do. Now, $H_t$ includes all healed since the start of the epidemic, hence a fraction of $H_t$ can be seronegative, depending on the duration $d$ of seropositivity. Hence we should compare 2.5\% to $H_t - H_{t-d}$, where $H_{t-d}=0$ if $t-d<1$. The true values of $H_t$ are unknown. We may recover them from $\hat{R}_t$ and the available data on the fraction of deaths as $\hat{H}_t = \hat{R}_t - D_t/u_{0.5}$, where $u_{0.5}$ is the median of the distribution of the observation error. Since Istat's survey was carried out between 25 May and 15 July 2020, we compare 2.5\% to
\begin{equation}\label{eq:mediaHhat}
	\bar{H} = \frac{1}{52}\sum_{t=\text{25 May}}^{\text{15 July}} (\hat{H}_t - \hat{H}_{t-d}) \ .
\end{equation}
A plausible value of $d$ is three months (\cite{duysburgh2021persistence}). This procedure rests on several assumptions and we only regard it as a way to check for gross deviations of our model from reality.

\subsection{State and parameter estimation}
We run the RBPF algorithm with 20,000 particles and time discretisation step of $1/24$ day as done for the synthetic data.
The initial values are $(\beta_0, \gamma_0)^T=\mu_0=(0.3,0.1)^T$ and $\Sigma_0=diag(0.05,0.02)$. Moreover $\sigma=0.03$ and $\eta=0.01$. 

The true IFR in a population of interest can only be known at the end of the epidemic, and having tested all the population. Because at the early stage of the epidemic reverse-transcription PCR (RT-PCR) diagnostic testing is primarily limited to people with significant indications of and risk factors for COVID-19, and because a large number of infections with SARS-CoV-2 result in mild or even asymptomatic disease, the accurate estimation of IFR is challenging (\cite{mallapaty2020deadly}).
The Centre for Evidence-Based Medicine (CEBM) at the University of Oxford bases its timely updates of IFR point estimate on a continuously evolving meta-analysis based on CFR data. The IFR point estimate is then obtained by halving the lower bound of the 95$\%$ prediction interval of the CFR 
and the current estimate fixes the IFR at 0.54$\%$ 
(\cite{oke2020ifr}: last updated 2 February 2021; cebm.net/covid--19/global--covid--19--case--fatality--rates/).  In \cite{brazeau2020report} low and high income countries are separately discussed and in a typical high income country, with a greater concentration of elderly individuals, an overall IFR of 1.15$\%$ (0.78-1.79 95$\%$ prediction interval) is estimated. An estimate of 1.3$\%$ has been obtained  using data from the closed population of passengers in the Diamond Princess cruise ship (\cite{russell2020estimating}). The meta--analysis carried out by \cite{ meyerowitz2020systematic} of published research data
on COVID-19 infection fatality rates with last search on 16/06/2020 indicates a point estimate of IFR of 0.68$\%$ (0.53$\%$--0.82$\%$) with high heterogeneity, and suggests that in many populations the IFR would be $>1\%$ if excess mortality was taken into account.

\begin{figure}
\centering
\includegraphics[scale=0.6]{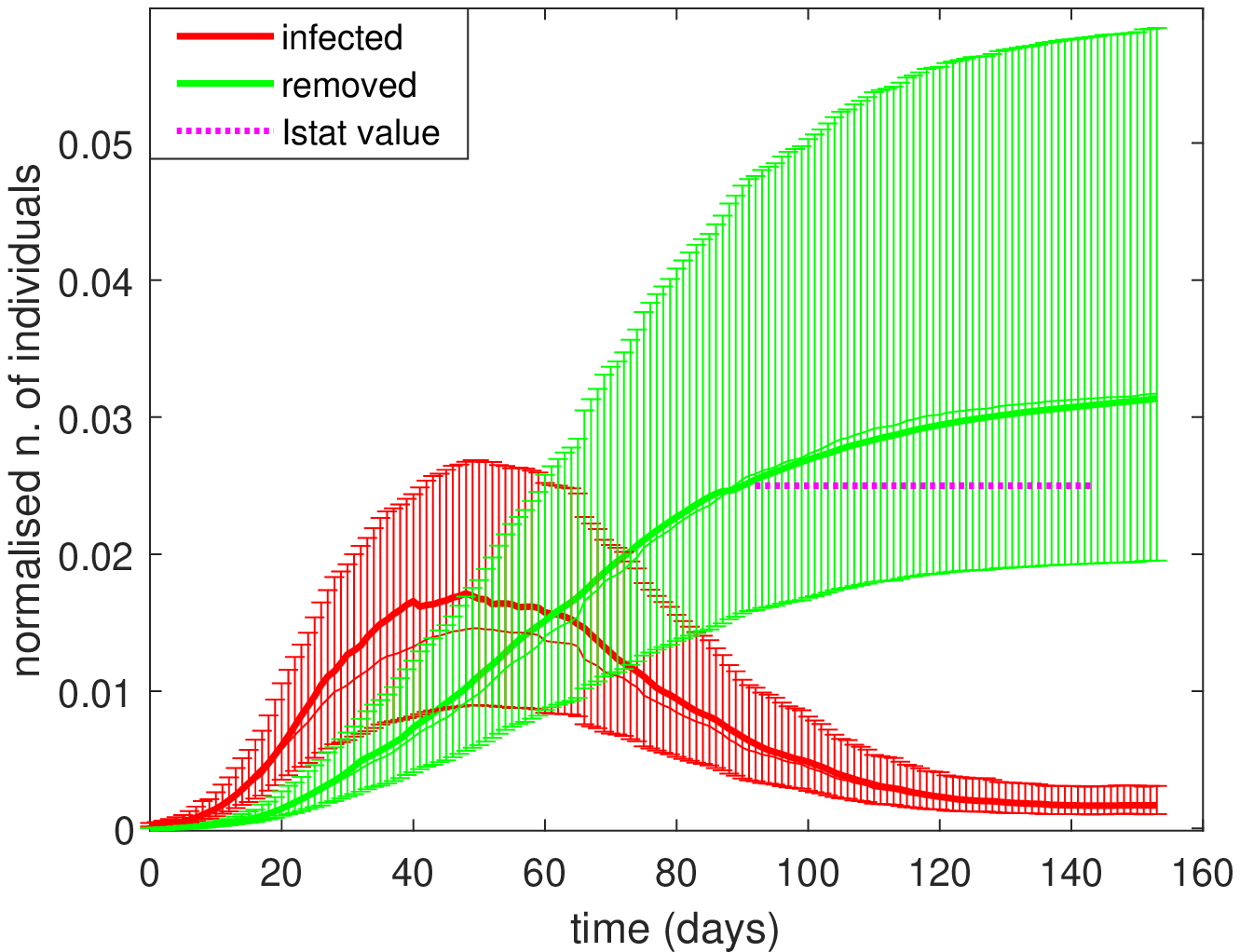}
\hspace{.5cm}
\includegraphics[scale=0.6]{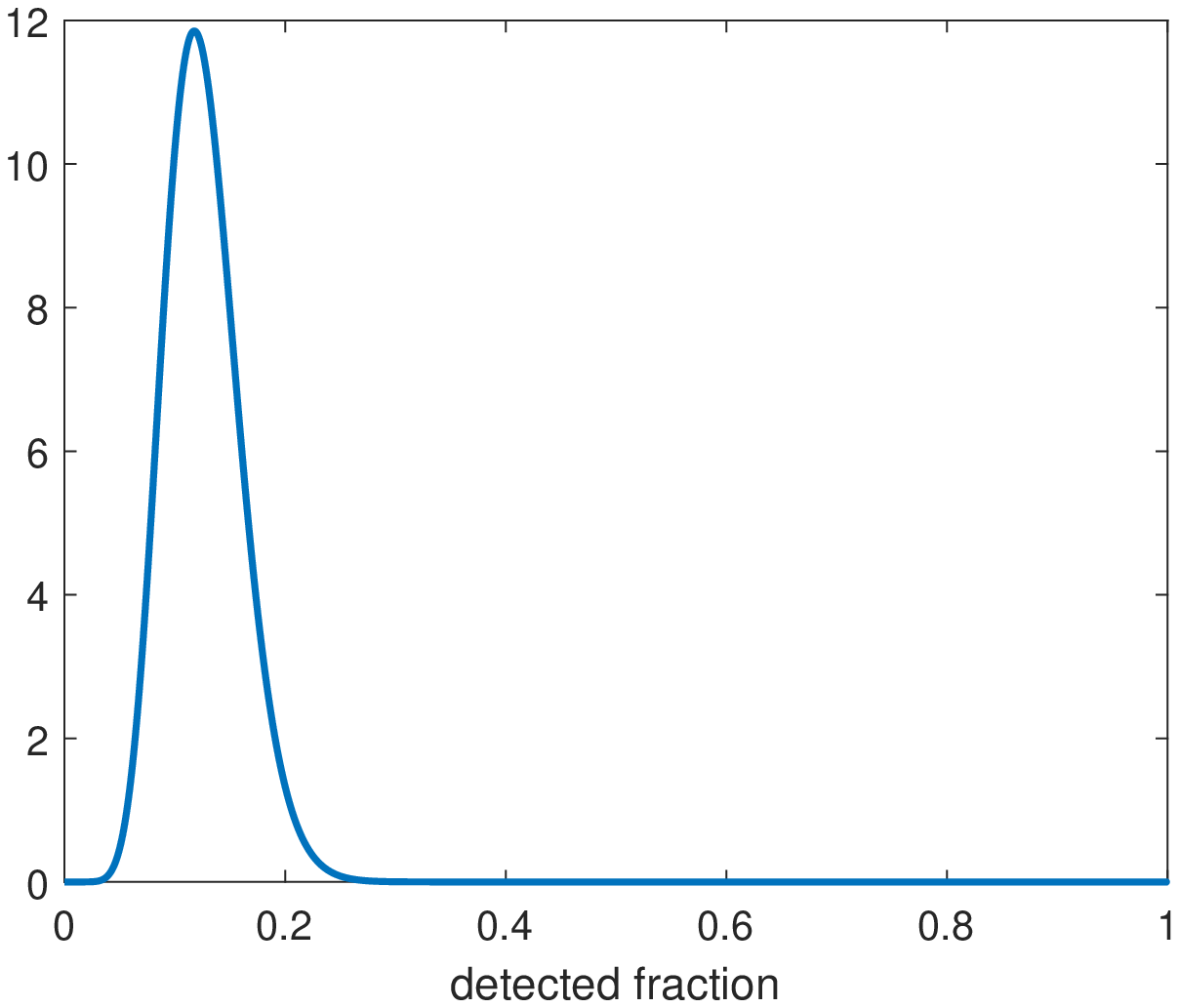}
\caption{Left panel: Filtered states for infected (thick red line) and removed (thick green line). Parameters of the beta observation error distribution (right panel) are $a=11.72$ and $b=81.39$ obtained considering IFR=$5\%$. 
 The prediction intervals are computed from \eqref{eq:cinf} with $q=0.025$. The thin lines are the observed infected (red) and removed (green) divided by $u_{0.5}$. The magenta dotted line represents the $2.5\%$ of the Italian population that have developed coronavirus antibodies in the Istat analysis from 25th May to 15th July. Time 0 is 1st March. }
\label{dinamiche_unicointervallo}
\end{figure}

For the first wave, we computed $a$ and $b$ from \eqref{eq:mm_rlambda} for a range of IFR values from 0.1$\%$--6$\%$, where the minimum value is the lowest we found in the relevant literature and has been suggested as lower bound of IFR in Europe by CEBM. The maximum value is still inspired by CEBM and by the considerations in \cite{ meyerowitz2020systematic}. Indeed, in Italy an estimated initial CFR of about 11-19$\%$ has been reported (\cite{de2020covid}, \cite{tosi2020clarification}, \cite{riccardo2020epidemiological}, \cite{CAVATAIO2021108545}). This suggested to consider as a possible maximum initial value an IFR=6$\%$,
accounting for the lack of knowledge at the beginning of the first wave. In particular, we considered $0.1\%$, $0.35\%$, $1.3\%$, and 6$\%$. Moving too far from the highest value gives a large discrepancy between Istat's 2.5\% estimated seropositivity in the population and \eqref{eq:mediaHhat}. Then we present results for $IFR=5\%$, for which $\bar{H} = 2.55\%$. 
Then $a=11.72$ and $b=81.39$ and the corresponding beta density is shown in the right panel of Figure \ref{dinamiche_unicointervallo}. The initial condition for $I_t$ ($R_t$), for each trajectory, is given by the normalized number of infected (removed) people collected by Protezione Civile on 1st  March divided by the median of this beta distribution.
The filtered states of infected and removed individuals are represented with thick lines in the left panel of Figure \ref{dinamiche_unicointervallo} where also the prediction intervals for both infected and removed are reported. The prediction intervals are computed from  \eqref{eq:cinf}.

\begin{figure}
\centering
\includegraphics[scale=0.8]{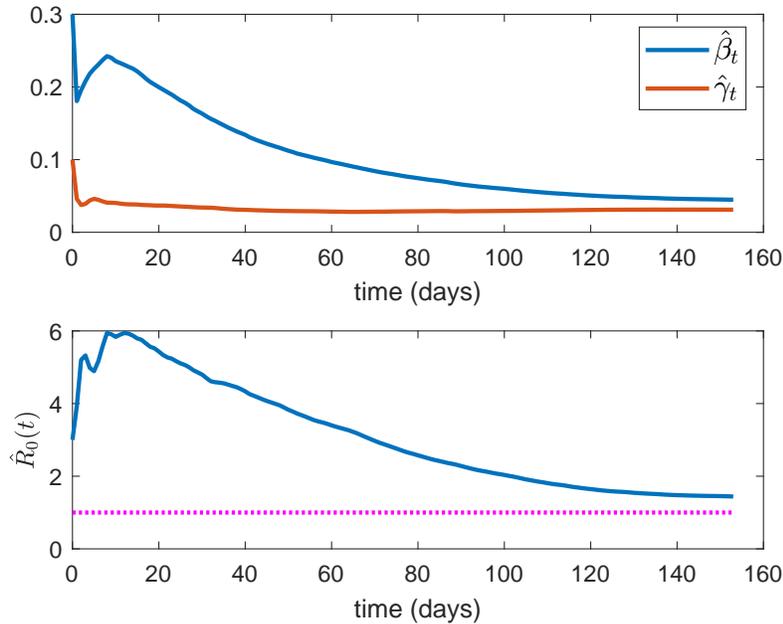}
\caption{Single time interval case. Top panel: plots of $\hat{\beta}_t$ and $\hat{\gamma}_t$ from \eqref{eq:meantheta}. Bottom panel: plot of $\hat{R}_0(t)$ from \eqref{eq:R0est}. Time 0 is 1st March.}
\label{parametri_unicointervallo}
\end{figure}

The plots of $\hat{\beta}_t$ and $\hat{\gamma}_t$ from \eqref{eq:meantheta} are in the top panel of Figure \ref{parametri_unicointervallo} and the plot of $\hat{R}_0(t)$ from \eqref{eq:R0est} is in the bottom panel.

The gap between the thick and the thin red lines in Figure \ref{dinamiche_unicointervallo} shows that a single SIR is not able to correctly describe the behaviour of the true dynamics. Therefore, we split the first wave into subintervals. For the partition we consider the DPCMs\footnote{DPCM: Italian acronym for government decrees. For a summary of the DPCMs related to the COVID-19 emergency see \url{http://www.governo.it/it/coronavirus-misure-del-governo}} with the greatest impact on social organization allowing for 10 days for the DPCM to have an effect on the epidemics (that is, change-points are the DPCM dates plus 10 days). In particular, we consider the following DPCM dates: 11th March, 22nd March, 26th April and 3rd June, so the change-points are on 21st March, 1st April, 3rd May and 13rd June.

\begin{figure}
\centering
\includegraphics[scale=0.8]{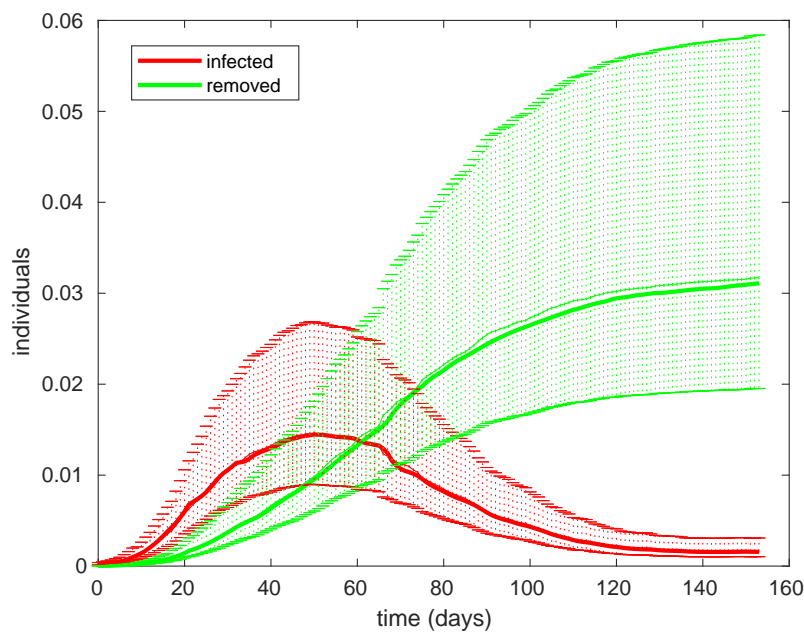}
\caption{Filtered states for infected (thick red line) and removed (thick green line) from five different SIRs
	 in the intervals $[0,20]$, $[20,31]$, $[31,66]$, $[66,104]$, $[104,160]$. Parameters of the beta observation error distribution are $a=11.72$ and $b=81.39$. The prediction intervals are computed from \eqref{eq:cinf} with $q=0.025$. The thin lines are the observed infected (red) and removed (green) divided by $u_{0.5}$. Time 0 is 1st March.}
\label{dinamiche_5intervalli}
\end{figure}

For each time interval, except for the first, we use the filtered state $\hat{x}_t$ \eqref{eq:filteredstate} at the end of the previous interval as initial state and the values of $\hat{\beta}_t$ and $\hat{\gamma}_t$ at the end of the previous interval as starting parameters. Then the discontinuity in the update is determined only by the initial covariance matrix. Since in the case of a single time interval we observed that the entries of $\Sigma_0$ are updated to very small values, then for the case of several intervals we do not restart each interval with $\Sigma_0=diag(0.05,0.02)$, but compromise between the updated $\Sigma_0$ at the end of the previous interval and the initial $\Sigma_0$, taking $\Sigma_0=diag(0.002,0.001)$, for all the time intervals. This choice also allows us to avoid big jumps in the trajectories of the parameters at the change-points.
The dynamics of the five different SIR models are represented as a whole dynamics in Figure \ref{dinamiche_5intervalli}.

\begin{figure}
\centering
\includegraphics[scale=0.8]{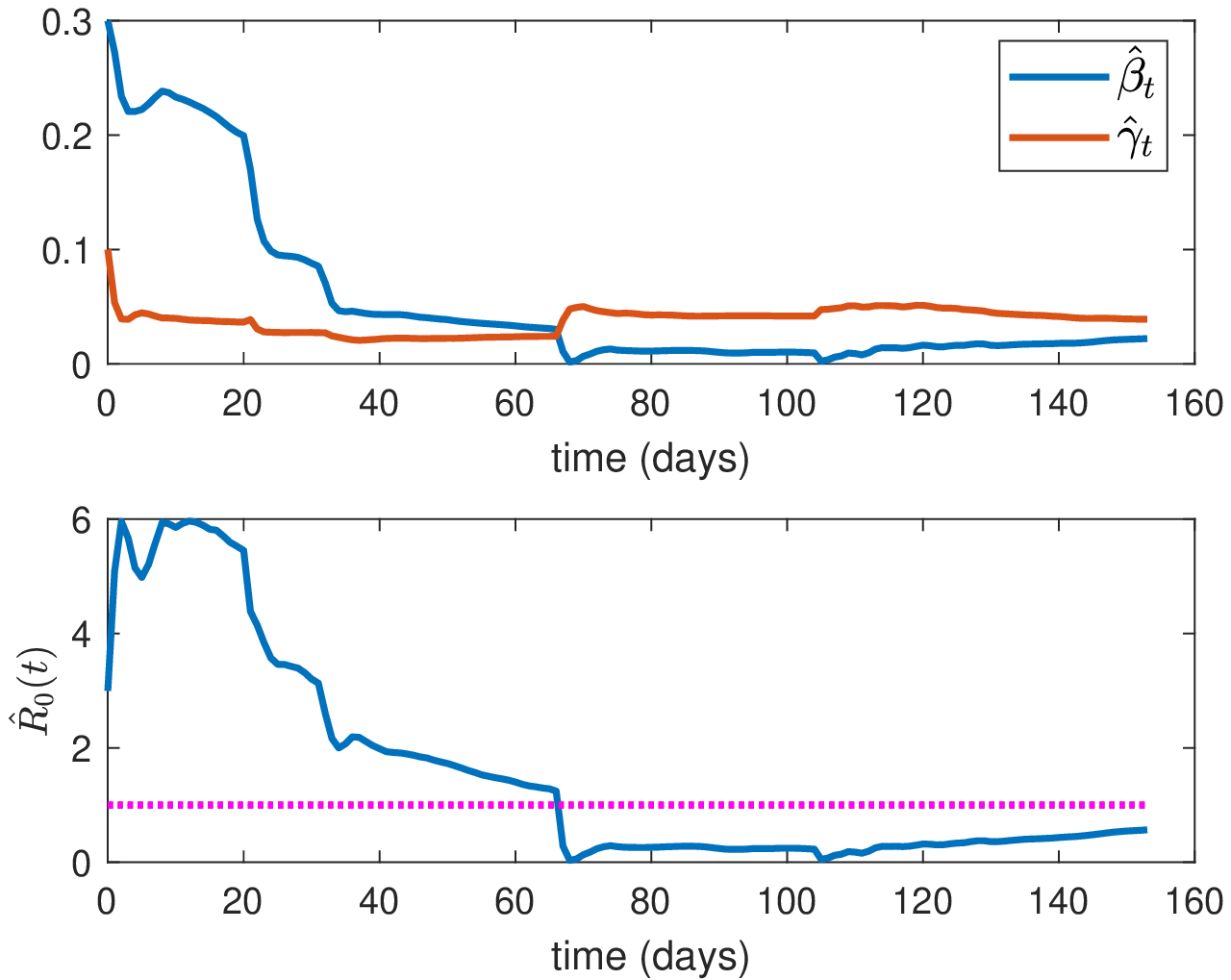}
\caption{Multiple time intervals case. Top panel: plots of $\hat{\beta}_t$ and $\hat{\gamma}_t$ from \eqref{eq:meantheta}. Bottom panel: plot of $\hat{R}_0(t)$ from \eqref{eq:R0est}. Time 0 is 1st March.}
\label{parametri_5intervalli}
\end{figure}

The trajectories $\hat{\beta}_t$, $\hat{\gamma}_t$ and $\hat{R}_0(t)$ show jumps at the change-points (Figure \ref{parametri_5intervalli}), not very pronounced due to the choice of a small $\Sigma_0$. After a few steps from each jump, the trajectories stabilize following a regular trend. The dynamics of infected individuals fits very well the observed infected divided by $u_{0.5}$.
After day 66 (corresponding to 26 April), $\hat{\beta}_t$ is smaller than $\hat{\gamma}_t$, so $\hat{R}_0(t)<1$. This value is more realistic than $\hat{R}_0(t)$ in the single time interval case, which is always greater than 1 (Figure \ref{parametri_unicointervallo}). 

This result is in agreement with the effective reproduction number published for the first time by Istituto Superiore di Sanit\`a (Italian National Istitute of Health, ISS) on 30 April (\cite{iss2020epidemia}): the effective reproduction numbers were reported for every Italian region (except for two because of bad quality data) and they were all smaller than one.

\begin{figure}
\centering
\subfigure{\includegraphics[scale=0.45]{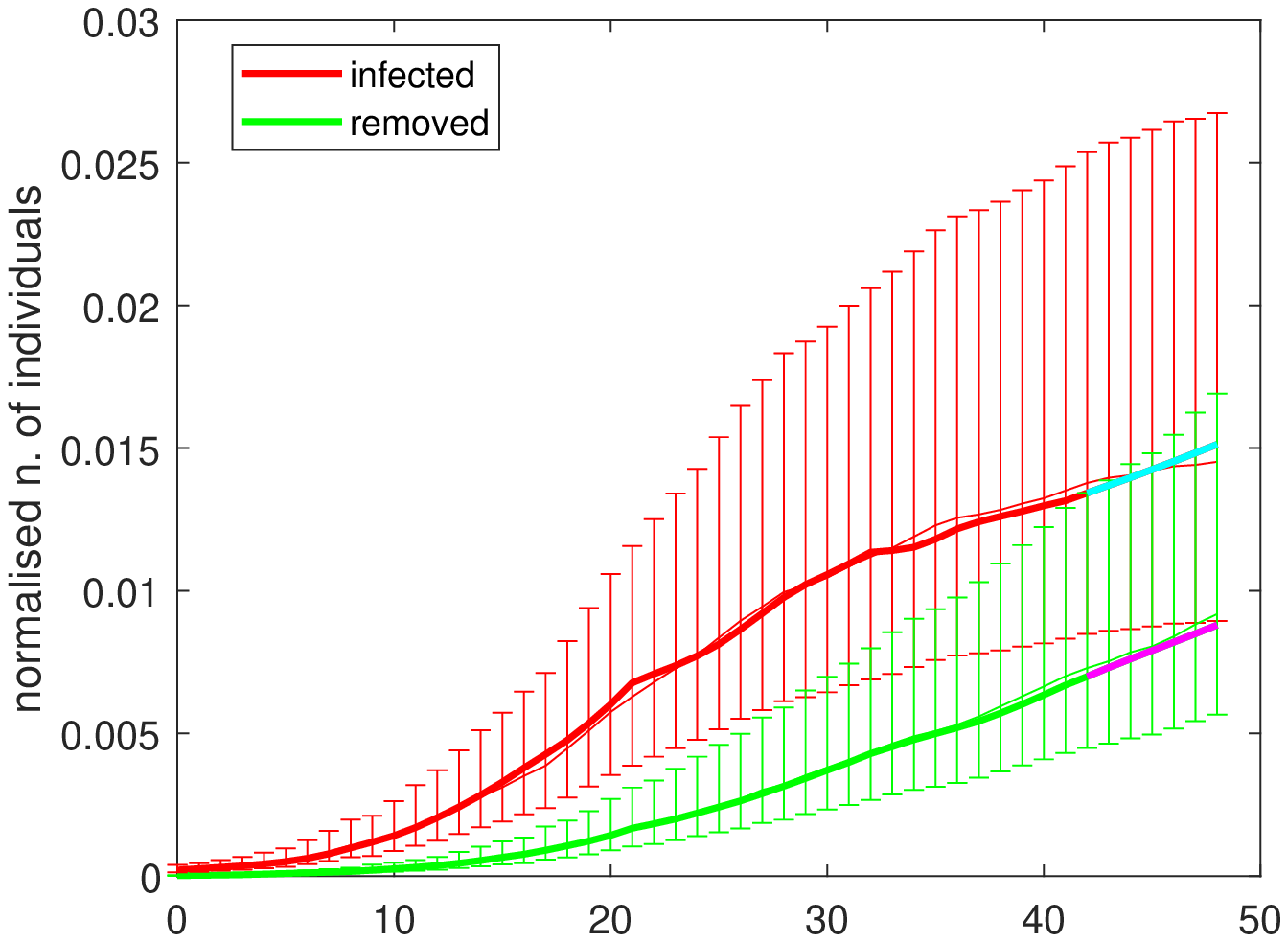}}
\hspace{.5cm}
\subfigure{\includegraphics[scale=0.45]{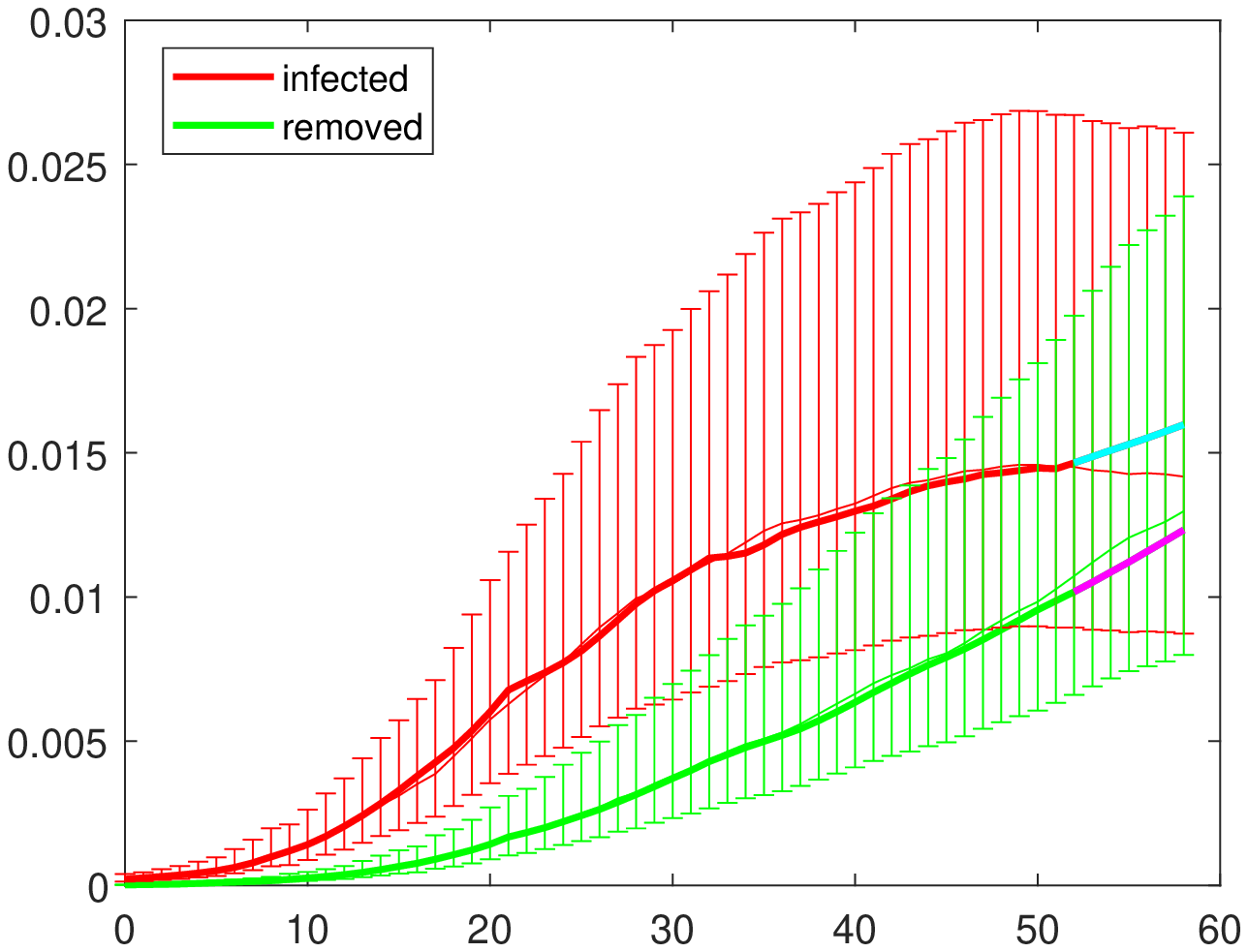}}
\hspace{.5cm}
\subfigure{\includegraphics[scale=0.45]{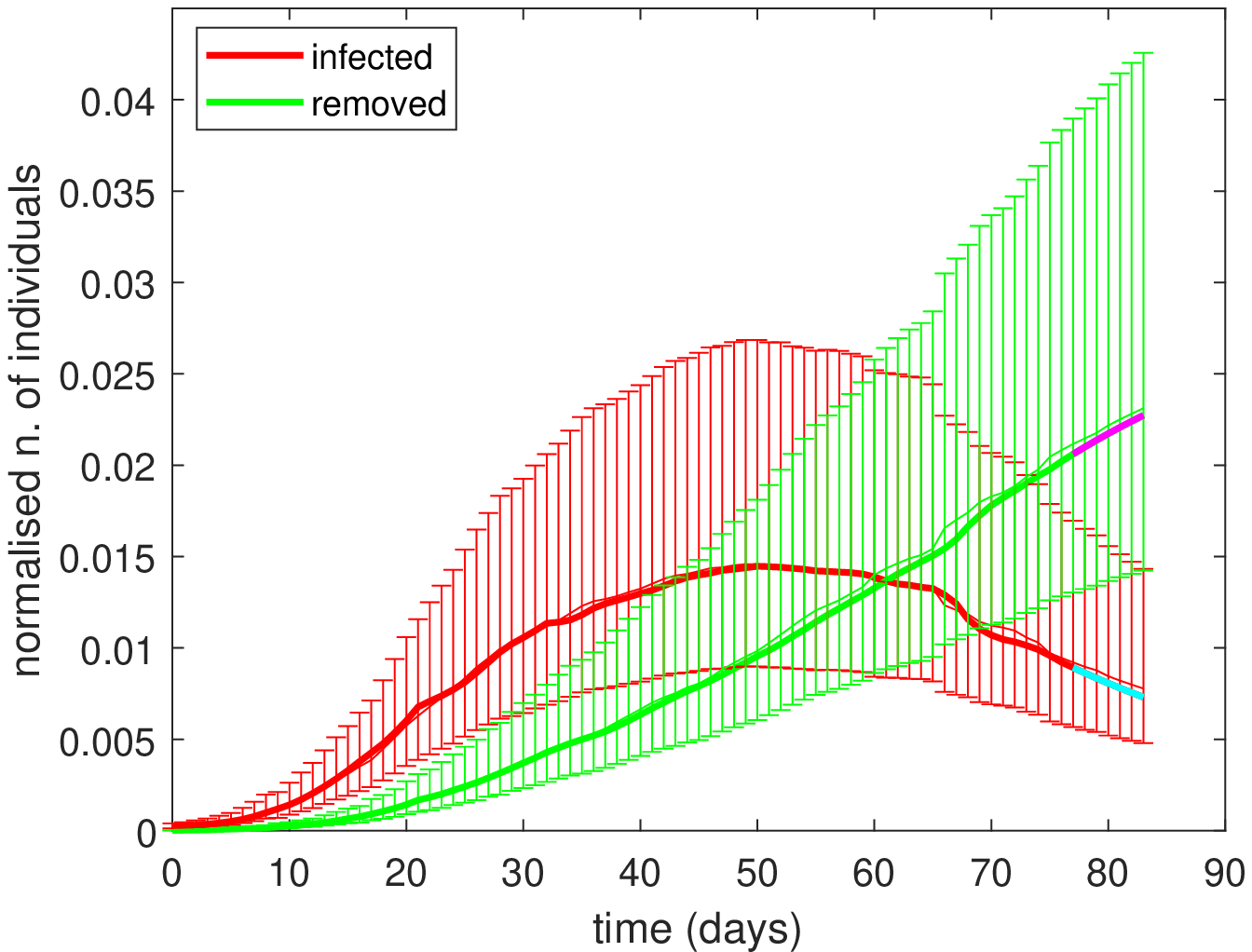}}
\hspace{.5cm}
\subfigure{\includegraphics[scale=0.45]{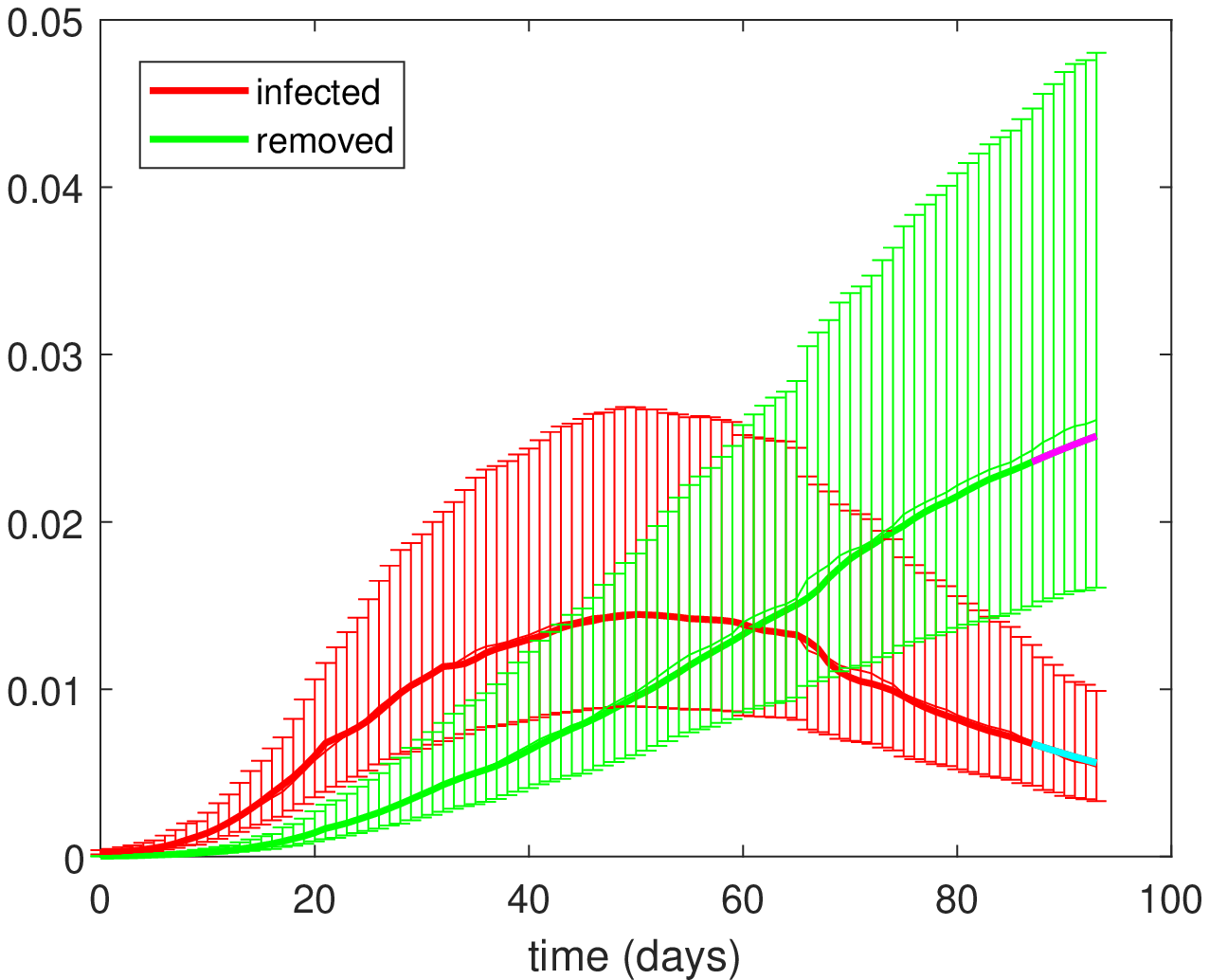}}
\caption{Dynamics of infected and removed individuals with forecasts during an increasing phase (first row) or a decreasing phase (second row). The forecasts start 10 days after a change point (first column) or 20 days after (second column). Starting dates are: 11th April (top left), 21st April (top right), 6th May (bottom left), 16th May (bottom right). Forecasts of infected and removed individuals are highlighted with different colours.}
\label{confronto_previsioni}
\end{figure}

Now, we analyze the forecast of the infected and removed dynamics for the first wave, computed as in \eqref{eq:forecastMC}. We consider different cases: we suppose to have observations up to 10 days or 20 days after the second or third change-points, and we try to forecast the dynamics for 7 future days (Figure \ref{confronto_previsioni}).

It can be observed from Figure \ref{confronto_previsioni} that the forecast is satisfactory when it starts 10 days after a change-point (left panels), while when it starts 20 days after a change-point is satisfactory only in the decreasing phase (bottom right panel). This different performance is mainly due to how fast $\hat{\beta}_t$ changes after each change-point, rather than to the number of observations taken before the forecast is started.

Now, we focus on the second wave, starting on 1st August and we select three change-points 10 days after the following measures: on 1st September (partial opening of museums, stadiums, and the increase in public transport occupancy); on 21st October (curfew in Lombardy, the most affected region); on 3rd November (DPCM establishing red, orange and yellow scenarios to classify the Regions from the highest to the lowest risk and introducing tiered restrictions). 
We run the RBPF algorithm, with $\mu_0=(0.05,0.03)^T$ as initial value for $(\beta_0,\gamma_0)^T$ and $\Sigma_0=diag(0.002,0.001)$. Moreover, $\sigma=0.03$ and $\eta=0.01$. 
The state is formed by the infected individuals and by the new removed individuals since 1st August, that is, the difference between the collected removed at each time and the removed on 31st July.

\begin{figure}
	\centering
	\includegraphics[scale=0.6]{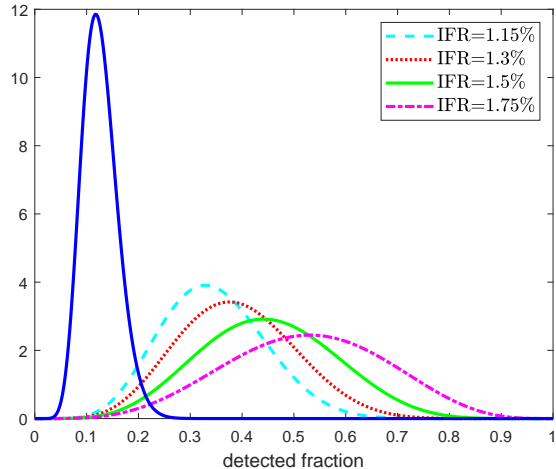}
	\caption{Beta distributions of the observation error for different values of IFR: $1.15\%$ (light blue dashed line), $1.3\%$ (red dotted line), $1.5\%$ (green continuous line), $1.75\%$ (magenta dashed-dotted line). For comparison also the beta density used for the first wave is reported (blue continuous line).}
	\label{confbeta}
\end{figure}

For the second wave the under-detection error of infected and removed people is smaller, because of an increase in resources for taking swab tests. 
For this reason it is appropriate to recalculate the parameters of the beta observation error distribution from \eqref{eq:mm_rlambda} only with the data since 1st August. Unfortunately, we lack a benchmark such as the serological survey during the first wave, and therefore we present the results obtained by considering four different $IFR$ values: $1.15\%$, $1.3\%$, $1.5\%$, $1.75\%$ according to the most recent studies. The beta densities obtained for these values are represented in Figure \ref{confbeta} and compared with the beta density used for the first wave. We exclude smaller values of $IFR$ because the corresponding observation error beta distributions are located on small values like for the first wave.  

\begin{figure}
\centering
\subfigure{\includegraphics[scale=0.5]{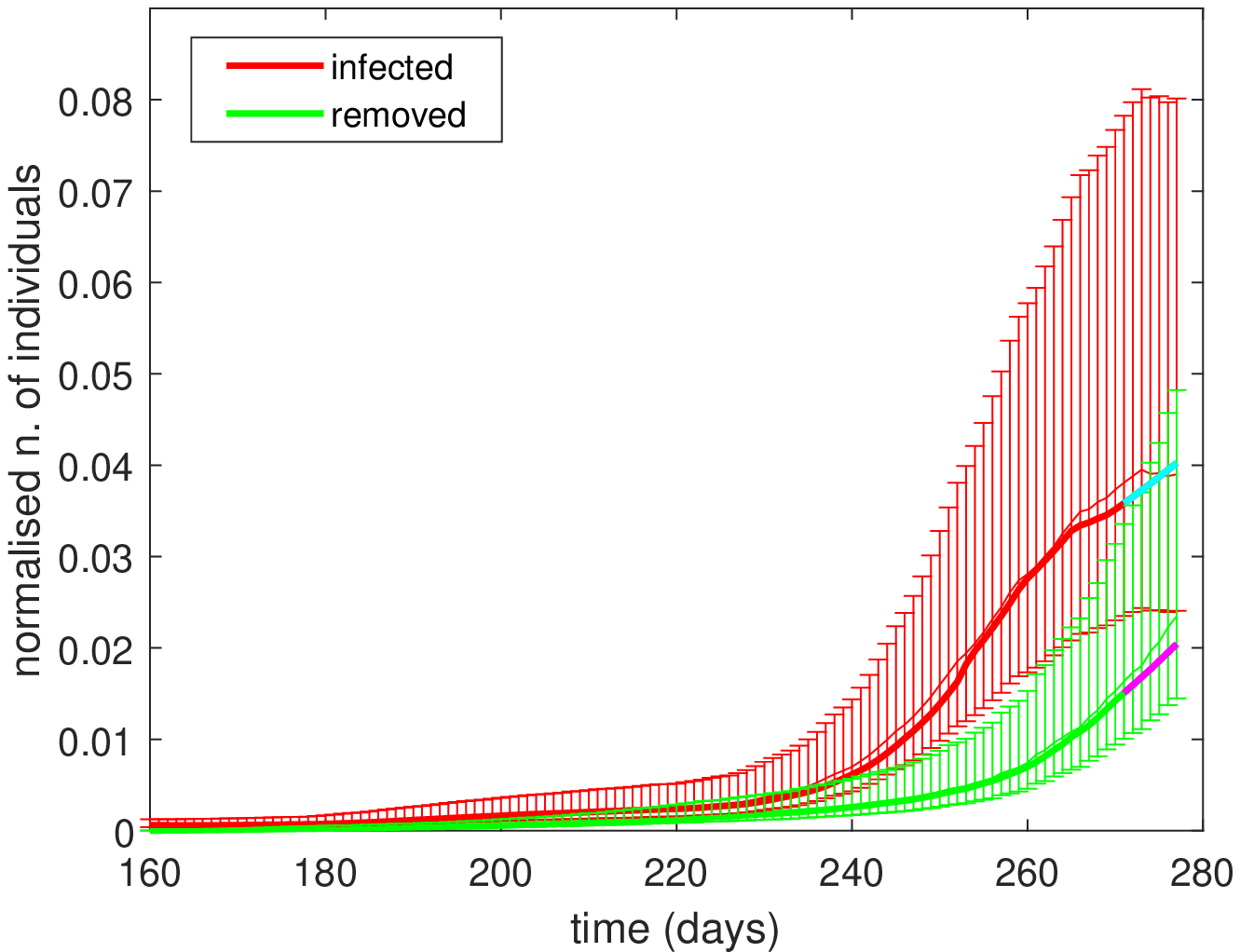}}
\hspace{.5cm}
\subfigure{\includegraphics[scale=0.5]{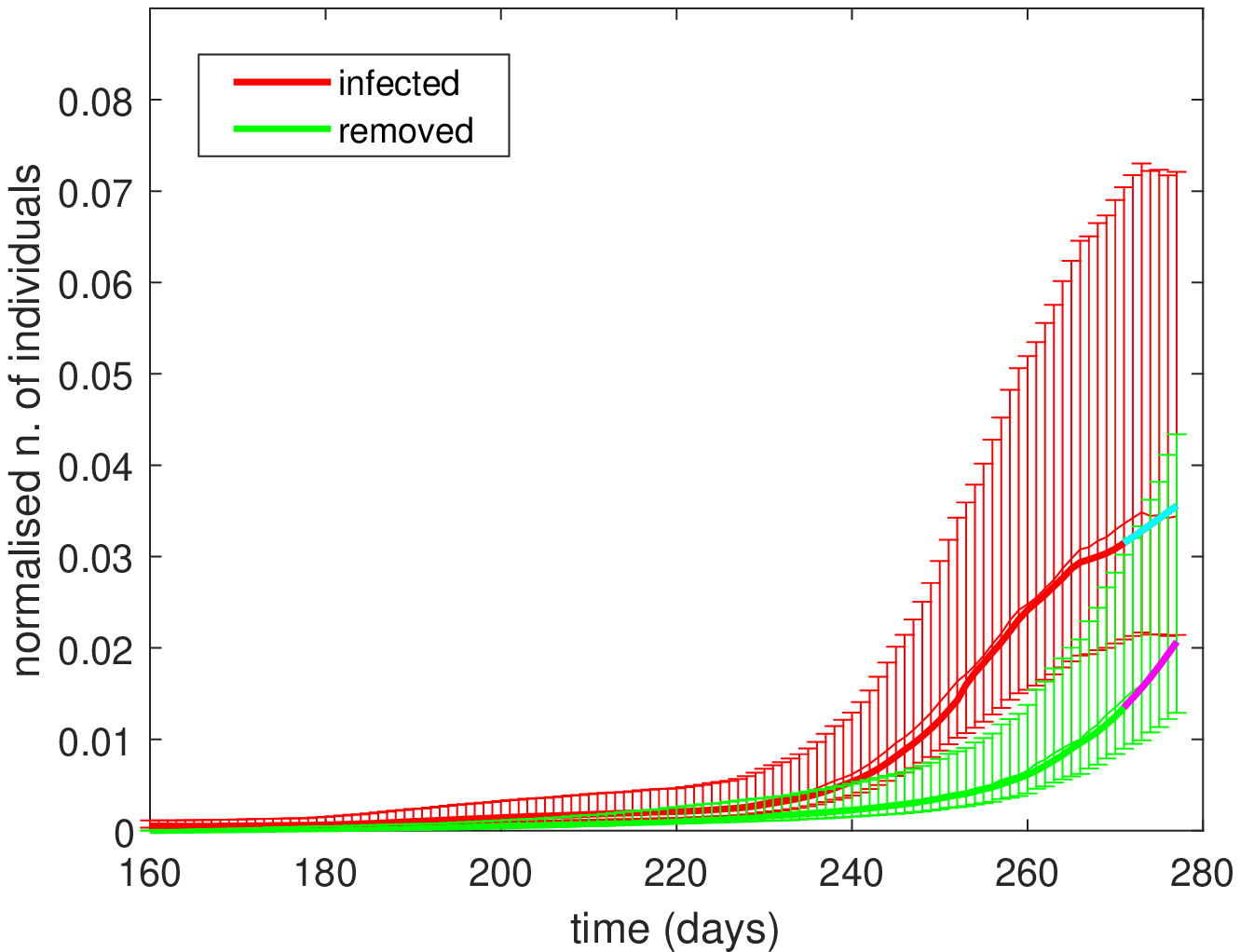}}
\hspace{.5cm}
\subfigure{\includegraphics[scale=0.5]{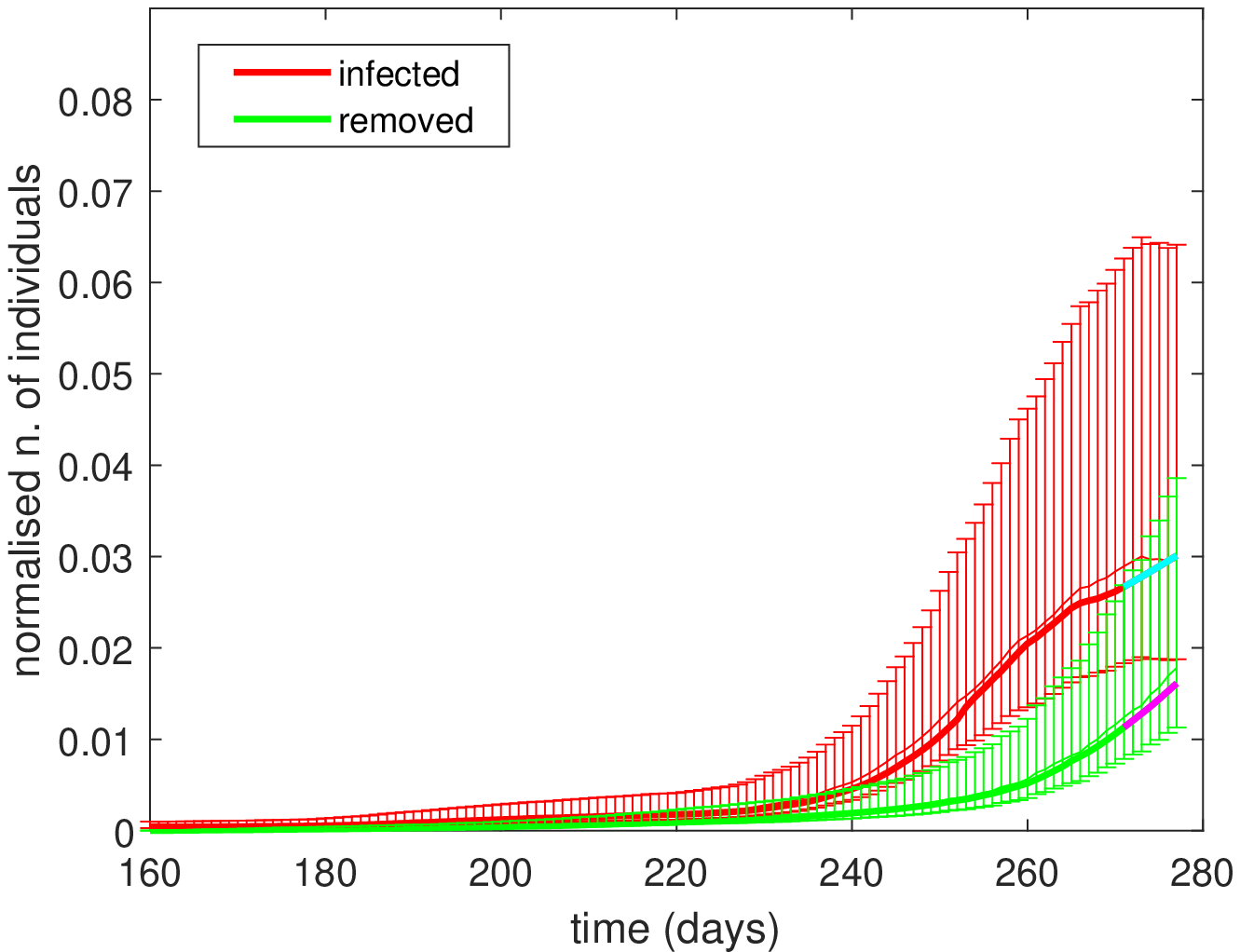}}
\hspace{.5cm}
\subfigure{\includegraphics[scale=0.5]{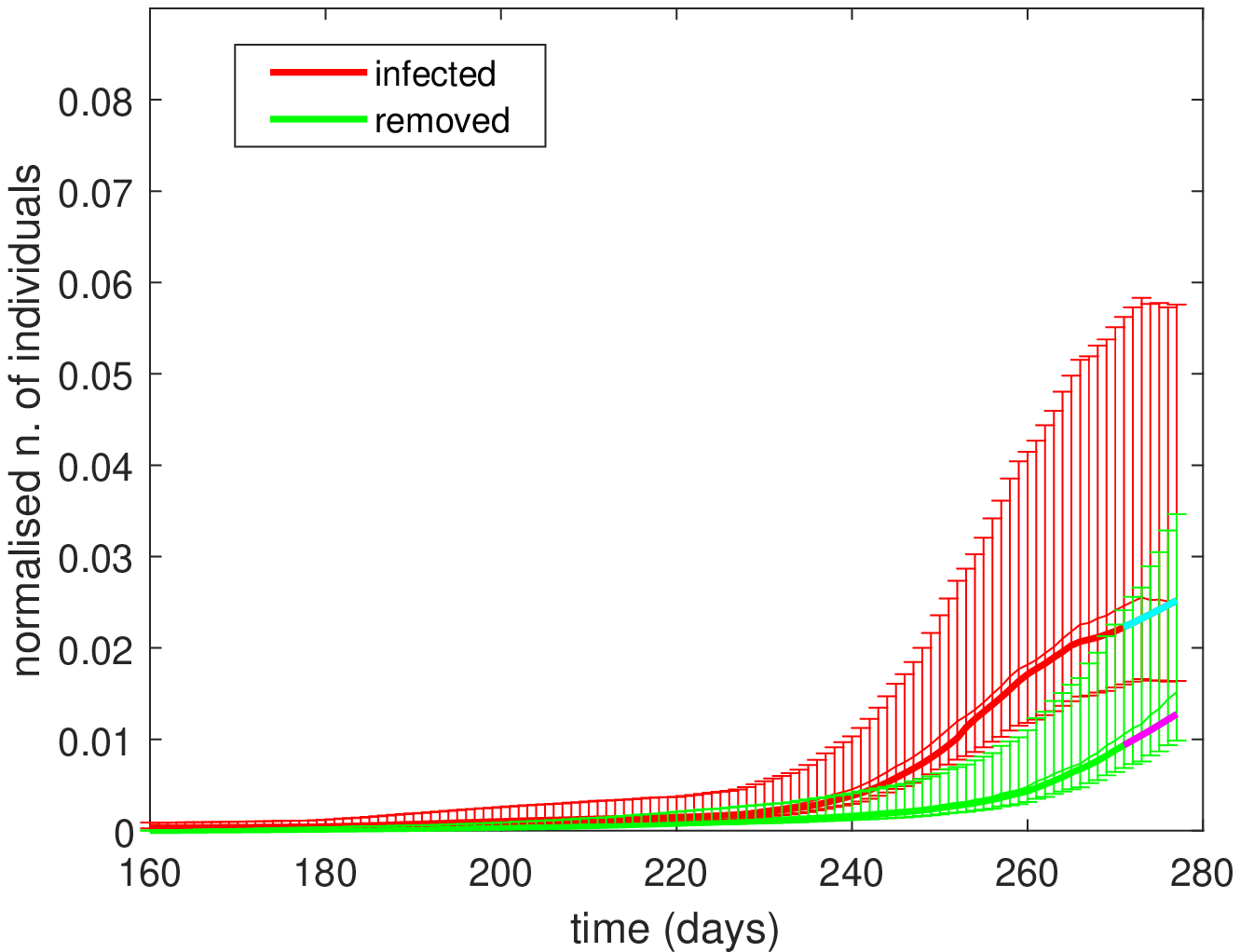}}
\caption{Filtered states for infected (thick red line) and removed (thick green line) in the second wave for different IFR: $1.15\%$ (top left), $1.3\%$ (top right), $1.5\%$ (bottom left), $1.75\%$ (bottom right).
 The prediction intervals are computed from \eqref{eq:cinf} with $q=0.025$. The thin lines are the observed infected (red) and removed (green) divided by $u_{0.5}$. Forecast of infected and removed individuals are highlighted with different colors. Time 160 is 1st August.}
\label{dinamiche_2ond}
\end{figure}

\begin{figure}
\centering
\includegraphics[scale=0.8]{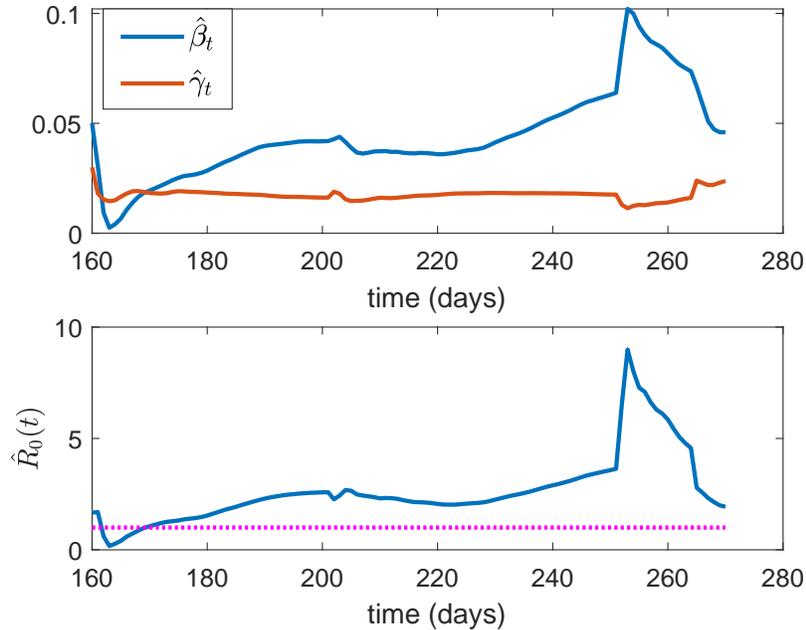}
\caption{Top panel: plots of $\hat{\beta}_t$ and $\hat{\gamma}_t$ from \eqref{eq:meantheta}. Bottom panel: plot of $\hat{R}_0(t)$ from \eqref{eq:R0est}. The IFR is $1.3\%$. Time 0 is 1st August.}
\label{parametri_2ond_IFR13}
\end{figure}

It can be seen from Figure \ref{confbeta} that both the mean and the variance of the beta distribution increase as the IFR increases. The corresponding dynamics are shown in Figure \ref{dinamiche_2ond}. The normalised numbers of individuals decrease when the IFR increases, because observations are divided by the median of the beta distribution which is increasing with the IFR. Also the width of the prediction intervals decreases as IFR increases. The filtered dynamics well reconstruct the behaviour of the adjusted data in all the cases. The estimated parameters $\hat{\beta}_t$ and $\hat{\gamma}_t$ are very similar for the different cases and only the parameters obtained using IFR=$1.3\%$ are reported in Figure \ref{parametri_2ond_IFR13}  with the corresponding  $\hat{R}_0(t)$. 

We observe a big jump in $\hat R_0(t)$ on the date of the second change point, 10 days after the curfew in Lombardy, followed by a slow decrease, denoting that this measure did not produce the desired effect. Then it was followed by the measure of 3rd November that allows $\hat R_0(t)$ to accelerate its decrease, approaching one, in agreement with what the ISS reported in its 25th November bulletin \cite{iss2020bepidemia}.

\subsection{A comparison with the SIR deterministic model}\label{sec:comparison}
\begin{figure}
	\centering
	\includegraphics[scale=0.5]{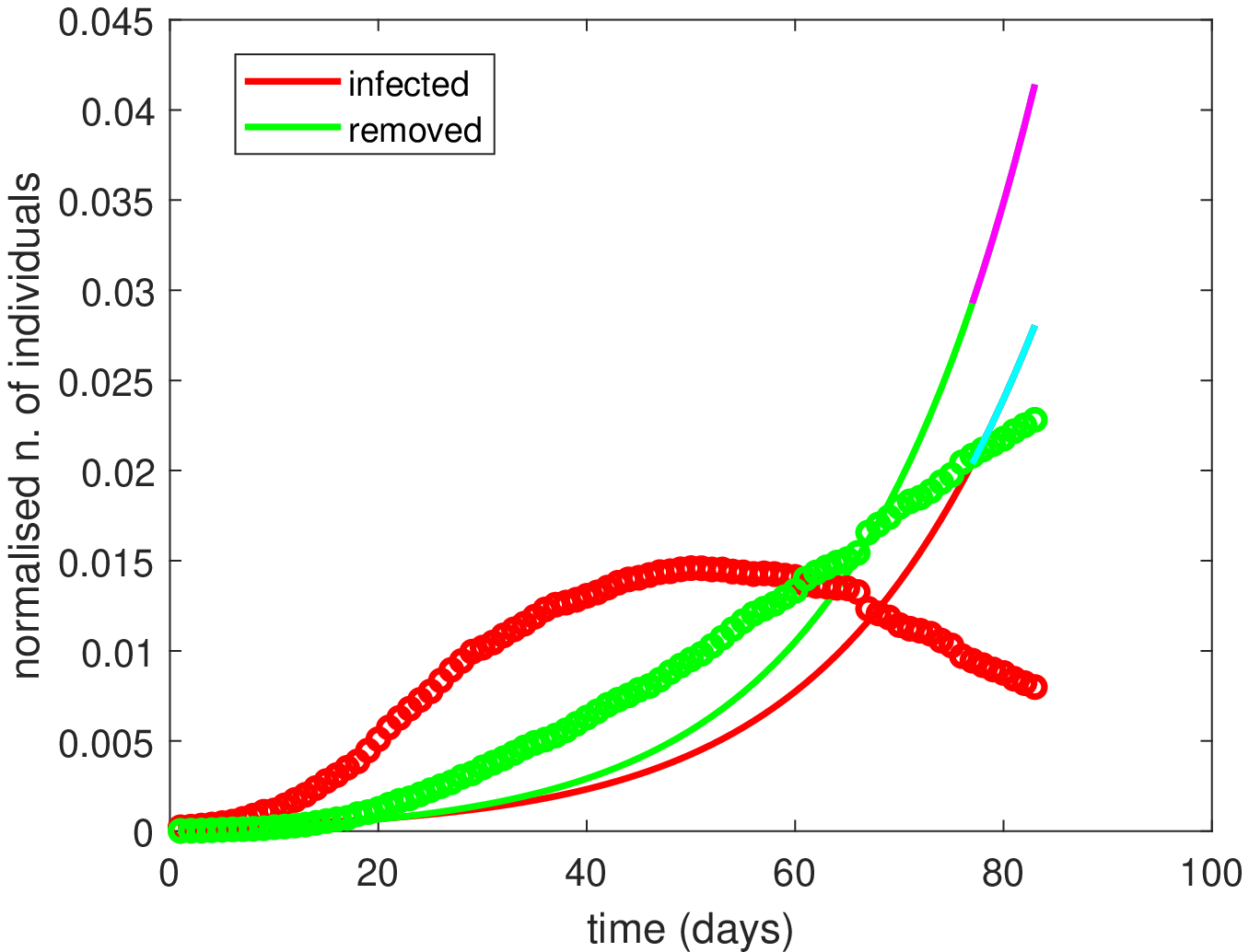}
	\hspace{.5cm}
	\includegraphics[scale=0.5]{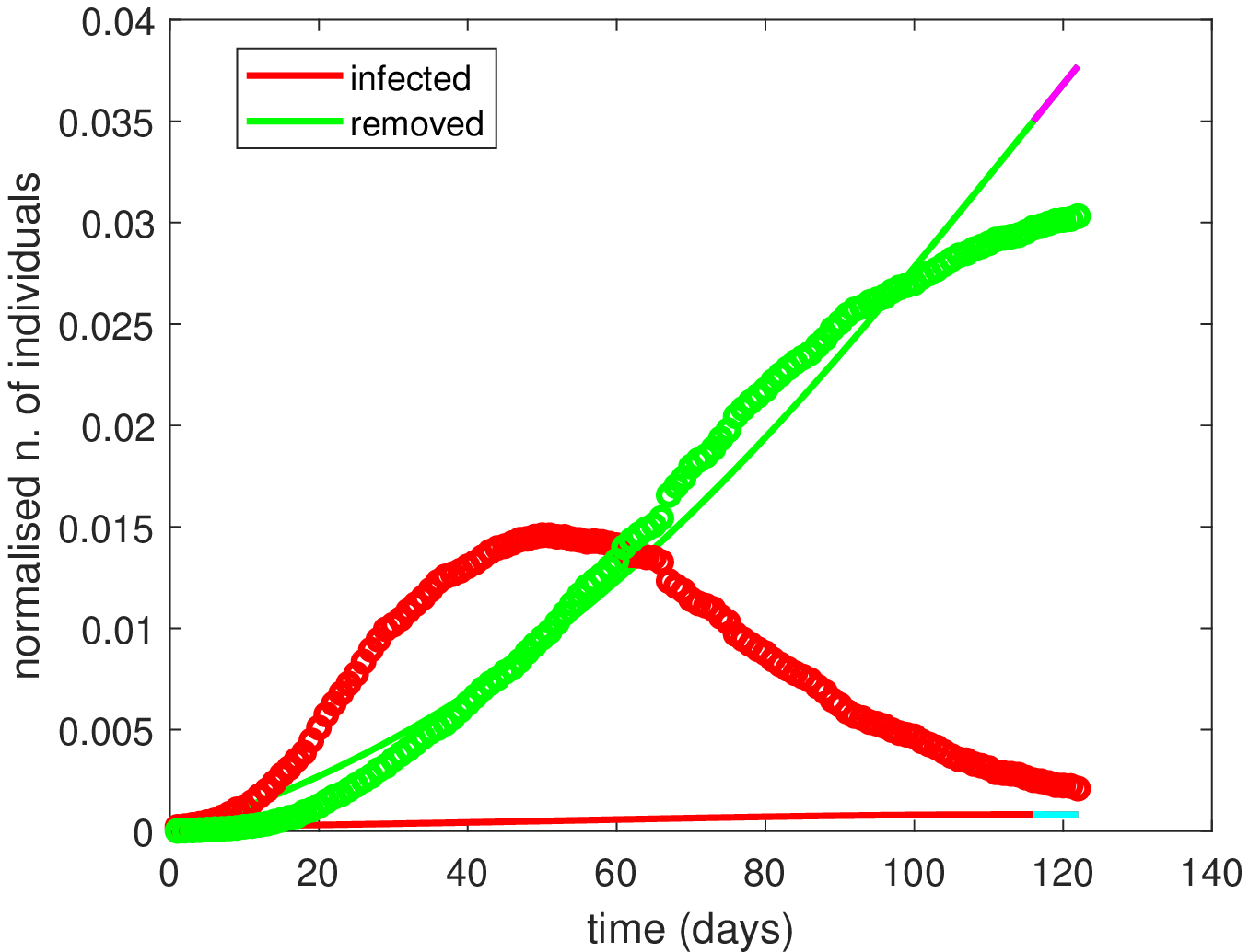}
	\caption{First wave: deterministic SIR simulations with ML parameter estimates and adjusted observations and seven-day forecast, until 15 May (left) and until 30 June (right). Estimated parameters are $(\beta,\gamma) = (0.15,0.08)$ and $(\beta,\gamma) = (0.57, 0.55)$,  respectively. Time 0 is 1st March.}
	\label{fig:dinam_sirdet_faseunica}
\end{figure}

\begin{figure}
	\centering
	\includegraphics[scale=0.5]{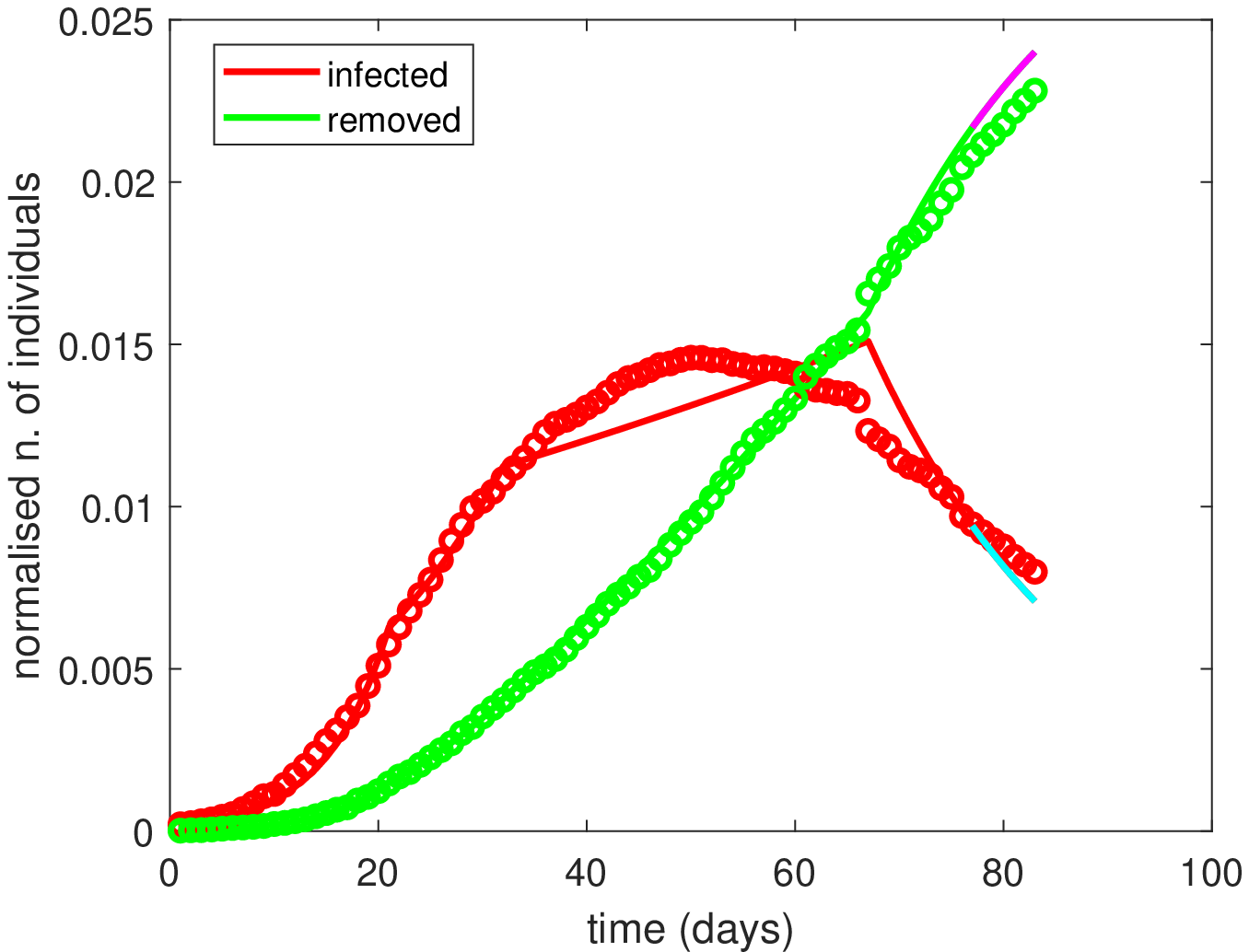}
	\hspace{.5cm}
	\includegraphics[scale=0.5]{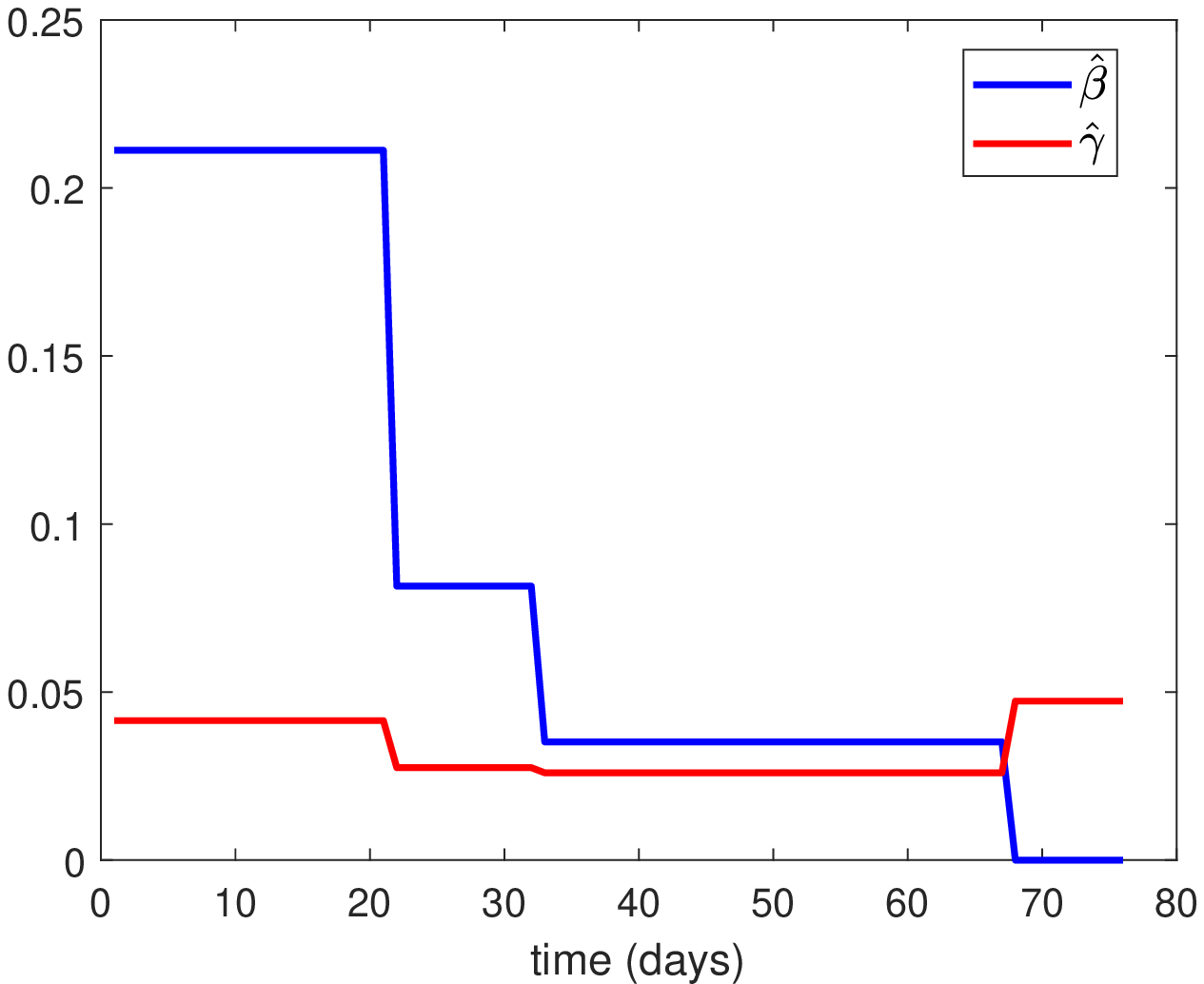}
	\caption{First wave: piecewise deterministic SIR simulations with ML parameter estimates and adjusted observations and seven-day forecast (left) until 15 May; parameter estimates $(\hat{\beta},\hat{\gamma})$ in each phase (right) are $(0.21,0.04)$, $(0.08,0.03)$, $(0.04,0.03)$, $(1.6\times 10^{-8},0.05)$. Cut dates correspond to 10 days after the government decrees on 11 March, 22 March, 26 April. Time 0 is 1st March.}
	\label{fig:dinam_sirdet_fasi}
\end{figure}

In this section we consider for comparison model \eqref{SIRdet} combined with the observation equations \eqref{errore_oss}, that is, the state dynamics is completely deterministic. We repeat part of the analysis done with the stochastic SIR model on the first wave data, using the same observation error distribution, and estimate $(\beta,\gamma)$ via maximum likelihood. In Figure \ref{fig:dinam_sirdet_faseunica} we show the single-phase deterministic SIR simulation  along with the adjusted observations, in two situations: when the number of infected is descending after the peak and when the descent is almost complete. In both cases the deterministic SIR is unable to capture the dynamics. The single-phase stochastic SIR model (see Figure \ref{dinamiche_unicointervallo}), although unsuitable, performs better thanks to the learning mechanism that makes adaptations both to the parameter and the filtered state. The piecewise deterministic SIR model, see Figure \ref{fig:dinam_sirdet_fasi}, follows the scaled observations more closely, still it is rather less flexible than its stochastic counterpart (see Figure \ref{dinamiche_5intervalli}). This is a further demonstration that a single SIR is unsuitable to describe the true behaviour of the pandemic.

\section{Concluding remarks}
In this work we have proposed a piecewise stochastic SIR model with change-points to the COVID-19 data in Italy from 24th February to 26th November 2020, using the dates of measures taken by the government to control the epidemic to define the change-points. The under-detection of the fractions of infected and removed in the population has also been explicitly modelled, introducing a distribution for the observation error. This strategy makes it possible to estimate the actual dynamics of the epidemic by overcoming the limitations of a single deterministic SIR model on the one hand and by correcting the observed data on the other hand. Then, via a particle filtering and parameter learning algorithm, the model can produce short-term predictions of the population in each compartment and continously updated estimates of key quantities such as the basic reproduction number, which can be acted upon by decision makers. We obtained a rather large basic reproduction number in the initial phase of the first wave, decreasing progressively in the following phases. This value may seem surprising, but other studies, such as \cite{linka2020reproduction}, confirm this behaviour which is common in European countries.

The stochastic SIR model might be used to evaluate the effect of mitigation measures by extending the predictions to a horizon of several weeks beyond the date of the next government decree, as an answer to the question: ``what would the mid-term evolution of the epidemics have been if this specific measure had not been taken''? But, given the complexity of the phenomenon, which is only partially captured by the model, a great deal of caution is required in doing so. A possible way out is to enrich the model by adding compartments, but, as our model identifiability study has shown, solid prior information or data relevant to the required additional parameters are needed to obtain meaningful results.

With respect to prior information, our approach to the selection of the observation error distribution depends on an estimate of the IFR. For the first wave we have chosen larger values than for the second wave (up to 6\% against up to 1.75\%), which is in agreement with a report of the ISS \cite{iss2021epidemia}, published after we finished our analyis, where the monthly standardised CFR has been calculated. When standardised with respect to the age and sex structure of the Italian population, the CFR is close to 9\% in February-March 2020 and close to 4.5\% in April. Then it falls to around 1\% in June and July and increases to above 2\% in October.


\section*{Acknowledgments}
The  authors  acknowledge  the  many  fruitful  discussions  with  several  colleagues,  and  in  particular  the colleagues at CNR-IMATI that participated in the COVID-19 modelling study group.

\appendix

\renewcommand{\theequation}{A-\arabic{equation}}
\setcounter{equation}{0}

\section{RBPF algorithm}\label{sec:RBPF}
To estimate parameter $\theta_0=\left(\beta_0,\gamma_0 \right)^T$ and state $X_t$ we propose to apply a Rao-Blackwellized particle filter (RBPF) algorithm.
We consider the Euler discretization of the stochastic system (\ref{SIR_vett}) reported in equation (\ref{SIR_disc}).
Since the system is linear in $\theta_0$, we can apply the Kalman filter. Suppose that $\theta_0=\left(\beta_0,\gamma_0\right)^T$ has a normal prior distribution with mean $\mu_0$ and covariance matrix $\Sigma_0$, then the distribution of $\theta_0$, given $x_{0:t+1} = \left(x_0,x_1,...,x_{t+1}\right)$ after $t+1$ time steps, as $t=0,1,2,\ldots$, is normal with mean $\mu_{t+1}$ and covariance matrix $\Sigma_{t+1}$ given by
\begin{equation}\label{kalman}
\begin{cases}
\mu_{t+1} & = \mu_{t}+S^T_{t+1}\left[x_{t+1}-x_{t}- h\left(x_t\right) \mu_t \Delta t \right]\\
\Sigma_{t+1} &= \Sigma_{t} - S^T_{t+1} h\left(x_t\right) \Sigma_{t} \Delta t \\
S^T_{t+1}&=\Sigma_{t} h^T\left(x_{t}\right) \Delta t
\left[h\left(x_t\right)\Sigma_{t} h^T\left(x_t\right) (\Delta t)^2 + g(x_t)g^T(x_t) \Delta t
 \right]^{-1}
\end{cases}
\end{equation}
The distribution of $X_{t+1}$ given $x_{0:t}$ is Gaussian with mean $B_{t+1}$ and covariance matrix $G_{t+1}$ given by
\begin{equation}
\begin{cases}\label{statepriorpars}
B_{t+1}&=x_t+ h \left(x_t\right) \mu_t \Delta t\\
G_{t+1}&=  h\left(x_t\right)\Sigma_{t} h^T\left(x_t\right) (\Delta t)^2 + g(x_t)g^T(x_t) \Delta t.
\end{cases}
\end{equation}

Recalling that the observations are obtained multiplying the state $X_t$ for the beta-distributed observation error term, as defined in equation (\ref{errore_oss}), the RBPF algorithm can be summarized as follows:

\noindent
STEP 1
\begin{itemize}
\item
At time $t=0$, we draw $M$ initial values of $X_0$ from its prior distribution $\pi\left(x_0\right)$ and obtain $M$ values $x_0^{(i)}, i=1,2,...,M$ or, alternatively, we put $x_0$ equal to the initial observation.
\item
We consider a prior distribution for the parameter $\theta_0$, given by a normal distribution ${\cal N} \left(\mu_0,\Sigma_0 \right)$, where $\mu_0$ is a vector of initial parameters, and $\Sigma_0$ is a diagonal covariance matrix.
\item
To obtain candidate values of the state at importance sampling steps, we will use the distribution implied by the state-transition equation \eqref{SIR_disc} after marginalising it with respect to $\theta_0$. At step one,  a value for $\tilde{X}_{1}^{(i)}$, conditional on $x_0^{(i)}$, is sampled from a normal distribution with mean $B_{1}^{(i)}$ and covariance matrix $G_{1}^{(i)}$, for $i=1,\ldots,M$, given by \eqref{statepriorpars} with $k=0$.
\item
Denoting by $y_1$ the observation at time $k=1$, we compute weights for each particle from the likelihood at $\tilde{x}_{1}^{(i)}$
$$
\tilde v_1^{(i)} = L(\tilde{x}_{1}^{(i)}; y_1) = p(y_{1,1} \vert \tilde{x}_{1,1}^{(i)}) \times p(y_{1,2} \vert \tilde{x}_{1,2}^{(i)})
$$
where  
\begin{equation}
p\left(y \vert x\right)=\frac{\left(\frac {y}{x} \right)^{a-1} \left(1-\frac {y}{x} \right)^{b-1}}{B\left(a,b\right)} \frac {1}{x} I_{[0,x]}(y).
\label{betacond}
\end{equation}

In order to resample the particles, we need to normalize the weights:
$$
v_{1}^{(i)}=\frac{\tilde{v}_{1}^{(i)}}{\sum_{i=1}^M\tilde{v}_{1}^{(i)}}.
$$

\item
We update the posterior distribution of $\theta_0$ given $\left\{\tilde{x}_{1}^{(i)},x_{0}^{(i)}\right\}$ by taking one step of the Kalman filter of equation \eqref{kalman}, obtaining the new mean vector $\tilde{\mu}_1^{(i)}$ and covariance matrix $\tilde{\Sigma}_1^{(i)}$.

\item
We resample $M$ particles from a discrete distribution with support $\left\{\left(\tilde{x}^{(i)}_{1}, \tilde{\mu}_1^{(i)}, \tilde{\Sigma}_1^{(i)}\right)\right\}_{i=1,\dots,M}$ and corresponding probabilities $\left\{v_{1}^{(i)}\right\}_{i=1,\dots,M}$. We denote by $\left\{\left(x^{(i)}_{1}, \mu_1^{(i)}, \Sigma_1^{(i)}\right)\right\}_{i=1,\dots,M}$ the resampled particles.
\end{itemize}

At time $t\ge 1$, assume the sample $\left\{\left(x^{(i)}_{t}, \mu_t^{(i)}, \Sigma_t^{(i)}\right)\right\}_{i=1,\dots,M}$ is available.

\bigskip
STEP $t+1$
\begin{itemize}
\item
For $i=1,\ldots,M$, sample candidate particles $\tilde{x}_{t+1}^{(i)}$ from a normal distribution with mean $B_{t+1}^{(i)}$ and covariance matrix $G_{t+1}^{(i)}$,  given by \eqref{statepriorpars}.
\item
Compute the weights $\tilde v_{t+1}^{(i)}$ for each particle as the product of two distributions with density (\ref{betacond}). Normalize the weights:
$$
v_{t+1}^{(i)}=\frac{\tilde{v}_{t+1}^{(i)}}{\sum_{i=1}^M\tilde{v}_{t+1}^{(i)}}.
$$
\item
Update the posterior distribution of $\theta_0$ given $x_{0:t+1}^{(i)}$, which is a normal distribution with mean  $\tilde{\mu}_{t+1}^{(i)}$ and covariance matrix $\tilde{\Sigma}_{t+1}^{(i)}$ given by equation \eqref{kalman}.
\item
Resample $M$ particles using the probabilities $\left\{v_{t+1}^{(i)}\right\}_{i=1,\dots,M}$ and denote the resampled particles by $\left\{\left(x^{(i)}_{t+1}, \mu_{t+1}^{(i)}, \Sigma_{t+1}^{(i)}\right)\right\}_{i=1,\dots,M}$.
\end{itemize}

Particles $\left\{\left(x^{(i)}_{t}, \mu_{t}^{(i)}, \Sigma_{t}^{(i)}\right)\right\}_{i=1,\dots,M}$ are a sample from the distribution of interest. In detail,  the $x^{(i)}_{t}$'s are a sample from $p(x_t|y_{1:t})$ and, by keeping track of the resampling history, the entire sample $x_{0:t}^{(i)}$, $i=1,\ldots,M$ is potentially available, hence a sample from $p(x_{0:t}|y_{1:t})$. The mean of $x_t^{(i)}$ over the particles approximates $E(x_t|y_{1:t})$ and we call it the filtered state:
\begin{equation}\label{eq:filteredstate}
	\hat{x}_t = \frac{1}{M} \sum_{i=1}^M x_t^{(i)} \ .
\end{equation}
The $\mu_{t}^{(i)}$'s and $\Sigma_{t}^{(i)}$'s are a sample of conditional means and covariance matrices, that is,
$E(\theta_0|x_{0:t}^{(i)})$ and $Cov(\theta_0|x_{0:t}^{(i)})$. Therefore, an estimate of $E(\theta_0|y_{1:t})$ is
\begin{equation}\label{eq:meantheta}
(\hat{\beta}_t, \hat{\gamma}_t)^T = \hat{\theta}_t = \frac{1}{M}	\sum_{i=1}^M  \mu_{t}^{(i)}
\end{equation}
and, by sampling $M$ values from $M$ Gaussian distributions $N(\mu_{t}^{(i)}, \Sigma_{t}^{(i)})$, $i=1,\ldots,M$, we produce a sample $(\theta_t^{(1)},\ldots,\theta_t^{(M)})$ from $p(\theta_0|y_{1:t})$.

The basic reproduction number is defined as $R_0 = \beta_0/\gamma_0$, therefore an estimate based on $y_{1:t}$ is $E(\beta_0/\gamma_0|y_{1:t})$, which is computed as
\begin{equation}\label{eq:R0est}
\hat{R}_0(t) = \frac{1}{M}	\sum_{i=1}^M \frac{\beta_t^{(i)}}{\gamma_t^{(i)}}
\end{equation} 
where $(\beta_t^{(i)}, \gamma_t^{(i)})^T = \theta_t^{(i)}$. If the variances on the diagonal of $\Sigma_{t}^{(i)}$ are small, the additional sampling from the $N(\mu_{t}^{(i)}, \Sigma_{t}^{(i)})$ may be unnecessary and the following approximation might be used, corresponding to degenerate conditional distributions:
\begin{equation}\label{eq:R0estapprox}
	\hat{R}_0(t) = \frac{1}{M}	\sum_{i=1}^M \frac{\mu_{1,t}^{(i)}}{\mu_{2,t}^{(i)}} \ .
\end{equation}
$\hat{R}_0(t)$ can be regarded as our best estimate of the basic reproduction number in the light of the observed data, not to be confused with the net or effective reproduction number.
A very useful discussion on the meaning of $R_0$ is provided by \cite{delamater2019complexity}.

\bibliography{biblio}
\bibliographystyle{abbrv}
\end{document}